\documentclass[aps,twocolumn,pra,floatfix,showpacs,lengthcheck,citeautoscript,superscriptaddress,longbibliography]{revtex4-1}
\usepackage{amsmath}
\usepackage{amssymb}
\usepackage{graphicx}
\usepackage{bm}
\usepackage{float}
\usepackage{mathtools}
\usepackage{color,soul}
\usepackage{times}
\usepackage{upgreek}
\usepackage{MnSymbol}
\usepackage{verbatim}
\usepackage[bottom]{footmisc}
\usepackage{bbold}

\usepackage[dvipsnames,svgnames,table]{xcolor}
\usepackage{hyperref}
\usepackage{fancyhdr}
\usepackage[english]{babel}
\usepackage[caption=false]{subfig}

\hypersetup{
pdfstartview={FitH},
colorlinks=true,    
linkcolor=NavyBlue, 
citecolor=Blue,   
filecolor=NavyBlue, 
urlcolor=NavyBlue   
}

\newcommand{\bea}{\begin{eqnarray}}
\newcommand{\eea}{\end{eqnarray}}
\newcommand{\be}{\begin{equation}}
\newcommand{\ee}{\end{equation}}

\begin{document}
\title{Nonequilibrium edge transport in quantum Hall based Josephson junctions}

\author{Lucila {Peralta Gavensky}}
\affiliation{Centro At{\'{o}}mico Bariloche and Instituto Balseiro,
Comisi\'on Nacional de Energ\'{\i}a At\'omica (CNEA)- Universidad Nacional de Cuyo (UNCUYO), 8400 Bariloche, Argentina}
\affiliation{Instituto de Nanociencia y Nanotecnolog\'{i}a (INN-Bariloche), Consejo Nacional de Investigaciones Cient\'{\i}ficas y T\'ecnicas (CONICET), Argentina}

\author{Gonzalo Usaj}
\affiliation{Centro At{\'{o}}mico Bariloche and Instituto Balseiro,
Comisi\'on Nacional de Energ\'{\i}a At\'omica (CNEA)- Universidad Nacional de Cuyo (UNCUYO), 8400 Bariloche, Argentina}
\affiliation{Instituto de Nanociencia y Nanotecnolog\'{i}a (INN-Bariloche), Consejo Nacional de Investigaciones Cient\'{\i}ficas y T\'ecnicas (CONICET), Argentina}

\author{C. A. Balseiro}
\affiliation{Centro At{\'{o}}mico Bariloche and Instituto Balseiro,
Comisi\'on Nacional de Energ\'{\i}a At\'omica (CNEA)- Universidad Nacional de Cuyo (UNCUYO), 8400 Bariloche, Argentina}
\affiliation{Instituto de Nanociencia y Nanotecnolog\'{i}a (INN-Bariloche), Consejo Nacional de Investigaciones Cient\'{\i}ficas y T\'ecnicas (CONICET), Argentina}

\begin{abstract}
We study the transport properties of a voltage-biased Josephson junction where the BCS superconducting leads are coupled via the edges of a quantum Hall sample. In this scenario, an out of equilibrium Josephson current develops, which is numerically studied within the Floquet-Keldysh Green's function formalism. We particularly focus on the time-averaged current as a function of both the bias voltage and the magnetic flux threading the sample and analyze the resonant multiple Andreev reflection processes that lead to an enhancement of the quasiparticle transmission. We find that a full tomography of the dc current in the voltage-flux plane allows for a complete spectroscopy of the one-way edge modes and could be used as a hallmark of chiral edge mediated transport in these hybrid devices.
\end{abstract} 
\maketitle

\section{Introduction}
The possibility of marrying superconductivity with the quantum Hall (QH) effect has brought to the table a plethora of novel physical phenomena: from the emergence of Andreev edge states~\citep{Hoppe2000,Giazotto2005} and crossed Andreev conversion~\citep{Hou2016,Lee2017} to realizations of non-Abelian anyons~\citep{Lindner2012,Clarke2013} and chiral Majorana fermions~\citep{Qi2010,Chaudhary2020}. In a series of recent experiments~\citep{Wan2015,Amet2016,Guiducci2018,Seredinski2019,Zhi2019,Zhao2020,Bhandari2020}, superconducting correlations were successfully induced at the edges of integer quantum Hall samples, paving the way to a new generation of such promising hybrid devices.

Several theoretical works have studied the mechanisms by means of which an equilibrium supercurrent flow can be established in Josephson junctions bridged by one-way edge states~\citep{Ma1993,vanOstaay2011,Stone2011,Alavirad2018,PeraltaGavensky2020}. The insulating nature of the bulk of the sample and the breaking of time-reversal symmetry cause the transfer of Cooper pairs to be realized via hybrid electron-hole edge modes that propagate chirally along the perimeter of the Hall bar~\citep{Hoppe2000,Bhandari2020}. In this scenario, the Josephson current-phase relationship is expected to behave in a peculiar manner as a function of the flux variations in the Hall device. In particular, the Fraunhofer pattern---which reveals the behavior of the critical current as a function of the magnetic field threading the sample--- is theoretically predicted to present periodic oscillations as a function of the normal flux quantum $\Phi_0 =hc/e$ \citep{Ma1993,vanOstaay2011,Stone2011}, a clear hallmark of chiral edge mediated transport. 

\begin{figure}[t]
\includegraphics[width=\columnwidth]{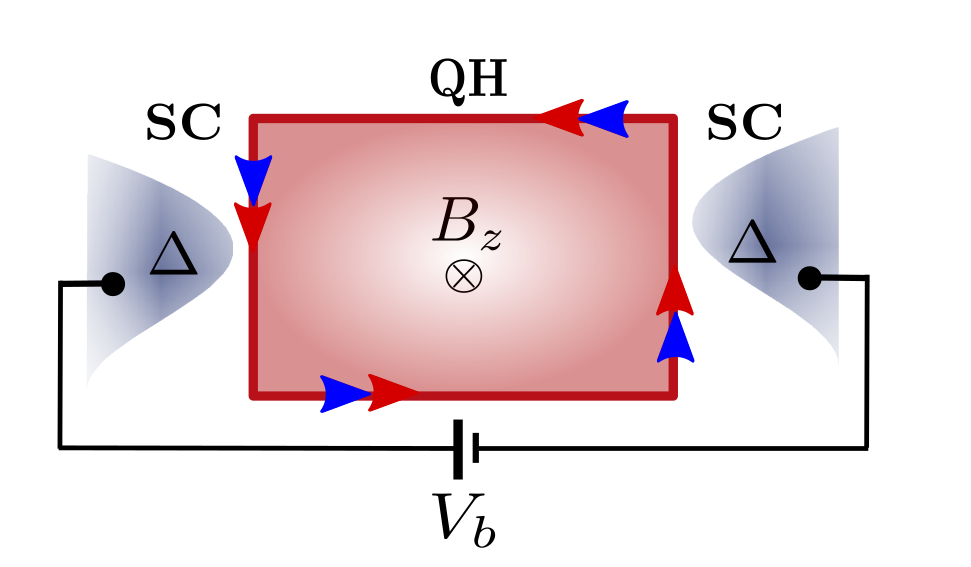}
\caption{Schematic setup: Two superconductors are coupled to a sample in the quantum Hall regime and a voltage bias $V_b$ is applied between them. The chiral nature of the edge states ensures that both the electrons and holes, depicted with arrows, flow with the same drift velocity along the perimeter of the Hall bar.}
\label{fig1}
\end{figure}
While much has been said about the equilibrium properties of these Hall based junctions, their response to inherently out-of-equilibrium transport experiments still remains a largely unexplored field. 
Our main aim in this paper is to partially fill this gap by analyzing the transport properties of a voltage driven superconductor (SC)-QH-SC Josephson junction such as the one depicted in Fig.~\ref{fig1}. The time-averaged current of voltage biased Josephson junctions is typically endowed with a rich subharmonic gap structure due to the presence of multiple Andreev reflection (MAR) processes that allow the transfer of quasi-particles from one terminal to the other~\cite{Klapwijk1983,Bratus1995,Averin1995,Cuevas1996}. These signatures in the current-voltage characteristic have been widely employed as a spectroscopy of the junction itself in a variety of superconducting devices involving single-level quantum dots~\cite{Yeyati1997,Johansson1999,Buitelaar2003}, molecules~\citep{Zazunov2006}, spin-split superconductors~\citep{Lu2020}, and even topological excitations~\cite{Meyer2011,SanJose2013,Huang2014}. 

In this paper, we theoretically investigate the current-voltage characteristic in our proposed QH setup as a function of the flux variations in the Hall bar and show how it could be used to unveil the presence of chiral edge mediated transport in the sample. We find that the time-averaged dc current of the device exhibits a series of distinctive resonances which are periodic with the superconducting flux quantum $\Phi_0^{s}=\Phi_0/2$, as opposed to the nondissipative equilibrium Josephson supercurrent. We interpret the appearance of this enhanced quasiparticle current within a Floquet multi-barrier picture of the resonant MAR processes. For voltages larger than the pairing gap, these peaks disperse linearly with the flux enclosed by the edge state due to resonant Cooper pair transfer between the superconducting leads, providing direct information on the drift velocity of the chiral channel. 

The paper is organized as follows. In Sec.~\ref{II} we introduce the low energy effective model which is used to describe the leads, the chiral edge state and their respective coupling. We include a discussion on the normal transmission of the junction, which allows for its characterization in terms of the microscopic parameters of the model. We also present the Floquet-Keldysh Green's function technique, which is employed to calculate the two-terminal out of equilibrium current when the terminals are superconducting. In Sec.~\ref{III} we present the numerically obtained current-voltage characteristic of this hybrid device and the interpretation of the results. In Sec.~\ref{IV} we analyze the main differences in the behavior of the time-averaged current when edge channels of both left and right chirality are present in the sample, modeling a typical Aharonov-Bohm configuration. In this setup, the chiral symmetry is broken when coupling the edge modes with the superconducting leads, allowing for backscattering to occur and thus changing drastically the flux dependence of the current-voltage characteristic. In Sec.~\ref{V} we summarize our main results and present some concluding remarks.
\section{Model and methods}\label{II}
\subsection{Hamiltonian approach}\label{IIA}
The model is depicted in Fig.~\ref{fig1}, where two BCS superconductors are coupled to a QH sample and a bias voltage $V_b = V_L - V_R$ is applied between them. We refer to $V_{R}$ and $V_{L}$ as the corresponding voltage of the right and left leads, respectively. 
Each superconducting terminal $\nu$ is modeled with a time-dependent Hamiltonian given by
\begin{equation}
\hat{H}_{\nu}\!=\!\sum_{\bm{k}\sigma}(\xi_{\bm{k}}-eV_{\nu})\hat{c}^{\dagger}_{\bm{k}\nu\sigma}\hat{c}^{}_{\bm{k}\nu\sigma}-\Delta\sum_{\bm{k}}\left(\hat{c}^{\dagger}_{\bm{k}\nu\uparrow}\hat{c}^{\dagger}_{\bm{-k}\nu\downarrow}e^{i\frac{2eV_{\nu}t}{\hbar}}+\text{H.c.}\right)\,,
\label{eqH}
\end{equation}
where $\xi_{\bm{k}}=\varepsilon_{\bm{k}}-\mu$ and $\Delta$ is the superconducting pairing amplitude, which is assumed to be the same in both leads. We have made use of the fact that the Cooper pair phase of each condensate acquires a time dependence due to the voltage drive, determined by the Josephson relation $\dot{\varphi}_{\nu}=2eV_{\nu}/\hbar$. 

We will focus on a regime where the magnetic field in the central region $\bm{B}=-B_z\,\bm{\hat{z}}$ is high enough to reach the extreme quantum limit. This being the case, the first Landau level is occupied and hence one chiral electronic and one hole like edge state of both spin species bridge both terminals. The central region is then described with a low-energy effective Hamiltonian as
\begin{eqnarray}
\notag
\hat{H}_{\text{ch}} &=& \sum_{\sigma}\int_{0}^{\Lambda}\hbar v_d \hat{\psi}_{\sigma}^{\dagger}(s)\left(-i\partial_s-\frac{2\pi}{\Phi_0}A\right)\hat{\psi}_{\sigma}^{}(s)\,ds\,,\\
&=&\sum_{n\sigma} \varepsilon_0\left(n - \frac{\Phi}{\Phi_0}\right)\hat{\psi}^{\dagger}_{n\sigma}\hat{\psi}^{}_{n\sigma}\,,
\label{Hch}
\end{eqnarray}
where $s$ is the coordinate along the edge of the sample, $-i\hbar\partial_s$ is its canonical momentum and the vector potential $\bm{A}=A\bm{\hat{s}}$ has been chosen in a gauge such that it remains parallel to the edge. The perimeter of the sample is identified with the variable $\Lambda$, and $\Phi_0 = hc/e$ is the normal flux quantum. The second equality is simply obtained by
expanding these fermionic fields in a plane wave basis $\hat{\psi}_{\sigma}(s) = \frac{1}{\sqrt{\Lambda}}\sum_n e^{i n \frac{2\pi s}{\Lambda}}\hat{\psi}_{n\sigma}$ imposing periodic boundary conditions such that $\hat{\psi}_{\sigma}(0)=\hat{\psi}_{\sigma}(\Lambda)$. Here $\varepsilon_0 = h v_d/\Lambda$ is the energy difference between consecutive levels of the isolated edge channel. The total flux enclosed by the edge state is given by $\Phi=\int_0^{\Lambda} \bm{A}\cdot\bm{ds}$. Within this model, $\Phi=0$ should be interpreted as a reference flux which is large enough to reach the quantum Hall limit and where a chiral edge Hall mode is pinned at the Fermi energy. In the simplified model described by Eq.~\eqref{Hch} we have neglected the Zeeman splitting between both occupied spin flavors. As we shall discuss in Sec. \ref{V}, taking into account this effect does not lead to significant changes in our main results.

The tunneling Hamiltonian that couples the leads with the Hall sample is given by
\begin{equation}
\hat{H}_T = -\gamma\sum_{\bm{k}\sigma}\left[\hat{\psi}_{\sigma}^{\dagger}(0)\hat{c}_{\bm{k}L\sigma} + \hat{\psi}_{\sigma}^{\dagger}(s_R)\hat{c}_{\bm{k}R\sigma} + \text{H.c.}\right]\,, 
\label{HT}
\end{equation}
where $\gamma$ is the tunneling amplitude and $s_R$ indicates the coordinate where the right superconducting terminal is attached. The total Hamiltonian is then given by $\hat{H} = \sum_{\nu}\hat{H}_{\nu} + \hat{H}_{ch} + \hat{H}_T$.
From hereon we use a symmetric bias such that $V_{L}=V_b/2$ and $V_{R}=-V_b/2$.

For technical reasons, it is convenient to perform a gauge transformation $\hat{U}(t)$ by means of which the time dependence of the leads is transferred to the tunneling matrix elements between the superconductors and the Hall sample. In particular, the time dependent part of the Hamiltonian can be solely included in the hopping to the left lead by taking the following transformation 
\begin{eqnarray}
\label{gauge}
\hat{U}(t)&=&\text{exp}\left[-i \sum_{\bm{k}\nu\sigma}\frac{e V_{\nu}t}{\hbar}c^{\dagger}_{\bm{k}\nu\sigma}c^{}_{\bm{k}\nu\sigma}\right]\\
\notag
& &\times\text{exp}\left[i\frac{e V_{b}t}{2\hbar}\sum_{\sigma}\int_{0}^{\Lambda} \hat{\psi}_{\sigma}^{\dagger}(s)\hat{\psi}^{}_{\sigma}(s) ds\right].
\end{eqnarray}
The total Hamiltonian transforms as 
$\hat{\widetilde{H}}=\hat{U}(t)\hat{H}\hat{U}^{\dagger}(t)-i\hbar\hat{U}\frac{d\hat{U}^{\dagger}}{dt}$, effectively generating the leads to behave as time independent
\begin{equation}
\hat{\widetilde{H}}_{\nu}=\sum_{\bm{k}\sigma}\xi_{\bm{k}}\hat{c}^{\dagger}_{\bm{k}\nu\sigma}\hat{c}^{}_{\bm{k}\nu\sigma}-\Delta\sum_{\bm{k}}\left(\hat{c}^{\dagger}_{\bm{k}\nu\uparrow}\hat{c}^{\dagger}_{\bm{-k}\nu\downarrow}+\text{H.c.}\right)\,,
\label{Hnu_g}
\end{equation}
and the coupling Hamiltonian as time dependent
\begin{equation}
\hat{\widetilde{H}}_T = -\gamma\sum_{\bm{k}\sigma}\left[\hat{\psi}_{\sigma}^{\dagger}(0)e^{i \frac{e V_b t}{\hbar}}\hat{c}_{\bm{k}L\sigma} + \hat{\psi}_{\sigma}^{\dagger}(s_R)\hat{c}_{\bm{k}R\sigma} + \text{H.c.}\,\right], 
\label{HT_g}
\end{equation}
while adding a diagonal energy to the chiral Hamiltonian
\begin{equation}
\hat{\widetilde{H}}_{\text{ch}} = \hat{H}_{\text{ch}} -\frac{e V_b}{2}\sum_{\sigma}\int_0^{\Lambda}\hat{\psi}^{\dagger}_{\sigma}(s)\hat{\psi}^{}_{\sigma}(s)\,ds\,.
\label{Hch_g}
\end{equation}
\subsection{Normal transmission of the model}
In this section we briefly discuss the transmission in the setup of Fig. 1 with normal leads ($\Delta=0$). This allows for a characterization of the transparency of the junction that will also prove to be useful when analyzing the transport properties when $\Delta\neq 0$. Since in the normal case we are dealing with a stationary situation (there is no time dependence arising from the  superconductivity), we will keep in this section the ungauged Hamiltonian defined by Eqs.~\eqref{eqH}-\eqref{HT}. Since we are interested in energy scales $e V_b \lesssim 2\Delta$, which are small compared with the variation of the leads' normal density of states, we take $\rho(\omega)=\rho(\varepsilon_F)$.

The transmission between the left and right lead can then be expressed in terms of the device Green's functions as~\citep{Datta1995}
\begin{equation}
T(\omega)= \Gamma_0 \mathcal{G}_{0\alpha}^{r}(\omega)\Gamma_{\alpha} \mathcal{G}_{\alpha 0}^{a}(\omega)\,,   
\label{T}
\end{equation}
where $\mathcal{G}_{0\alpha}^{r}$ ($\mathcal{G}_{\alpha 0}^{a}$) is the retarded (advanced) Green's function of the edge state that propagates from the site located at $s=0$ ($s_R= \Lambda\alpha/2\pi$) to the one at $s_R= \Lambda\alpha/2\pi$ ($s=0$). Here the angle $\alpha$ measures the position of the right lead relative to the left one and $\Gamma_0 = \Gamma_{\alpha} = 2\pi\rho(\epsilon_F)\gamma^2$. The non-local propagators in Eq.~\eqref{T} can be written in terms of the uncoupled Green's functions of the system by means of the equations of motion as
\begin{equation}
\mathcal{G}_{0\alpha}^{r}(\omega)=\frac{g_{0\alpha}^{r}(\omega)}{1-\gamma^2 \mathcal{D}_1(\omega)-\gamma^4 \mathcal{D}_2(\omega)}\,,   
\end{equation}
where
\begin{equation}
\mathcal{D}_1(\omega) = g^{r}_{00}(\omega)g^{r}_{LL}(\omega)+g^{r}_{\alpha\alpha}(\omega)g^{r}_{RR}(\omega)\,,
\end{equation}
and
\begin{equation}
\mathcal{D}_2(\omega)\!=\! g^{r}_{LL}(\omega)g^{r}_{RR}(\omega)[g^{r}_{0\alpha}(\omega)g^{r}_{\alpha 0}(\omega) - g^{r}_{00}(\omega)g^{r}_{\alpha\alpha}(\omega)]\,.
\end{equation}
In this limit, the retarded propagators of the left and right leads are simply given by $g_{LL}^{r} = g_{RR}^{r} = -i\pi\rho(\varepsilon_F)$. On the other hand, the required bare propagators of the central QH region may be obtained as
\begin{eqnarray}
g_{00}^{r}(\omega)=g_{\alpha\alpha}^{r}(\omega)&=&\frac{1}{\Lambda}\sum_n \frac{1}{\omega + i\eta -\varepsilon_0(n-\frac{\Phi}{\Phi_0})}\\
\notag
&=&\frac{\pi}{\Lambda\varepsilon_0}\cot\left(\pi\frac{\omega+i\eta}{\varepsilon_0}+\pi\frac{\Phi}{\Phi_0}\right)\, ,
\end{eqnarray}
while the non-local ones are
\begin{eqnarray}
\label{g0a}
g_{0\alpha}^{r}(\omega)&=&\frac{1}{\Lambda}\sum_n \frac{e^{-in\alpha}}{\omega + i\eta -\varepsilon_0(n-\frac{\Phi}{\Phi_0})}\\
\notag
&=&\frac{\pi}{\Lambda\varepsilon_0}e^{i(\pi-\alpha)\left(\frac{\omega+i\eta}{\varepsilon_0}+\frac{\Phi}{\Phi_0}\right)}\csc\left(\pi\frac{\omega+i\eta}{\varepsilon_0}+\pi\frac{\Phi}{\Phi_0}\right).
\end{eqnarray}
The Green's function $g_{\alpha 0}^{r}(\omega)$ is obtained by changing $\alpha\rightarrow 2\pi-\alpha$ in Eq.~\eqref{g0a}. Replacing these expressions in Eq.~\eqref{T} we finally obtain that the normal transmission between the leads in terms of the microscopic parameters of the model is given by
\begin{equation}
     T(\omega)=\frac{8 \tilde{\varepsilon}_0^2}{8 \tilde{\varepsilon}_0^2+(\tilde{\varepsilon}_0^2-1)^2\left\{1-\cos\left[2\pi\left(\frac{\Phi}{\Phi_0} +\frac{\omega}{\varepsilon_0}\right)\right]\right\}}\,,
\end{equation}
where we have defined $\tilde{\varepsilon}_0=\varepsilon_0/\pi \Gamma$ with $\Gamma = \pi\rho(\epsilon_F)\gamma^2/\Lambda$. 
\begin{figure}[t]
\includegraphics[width=0.85\columnwidth]{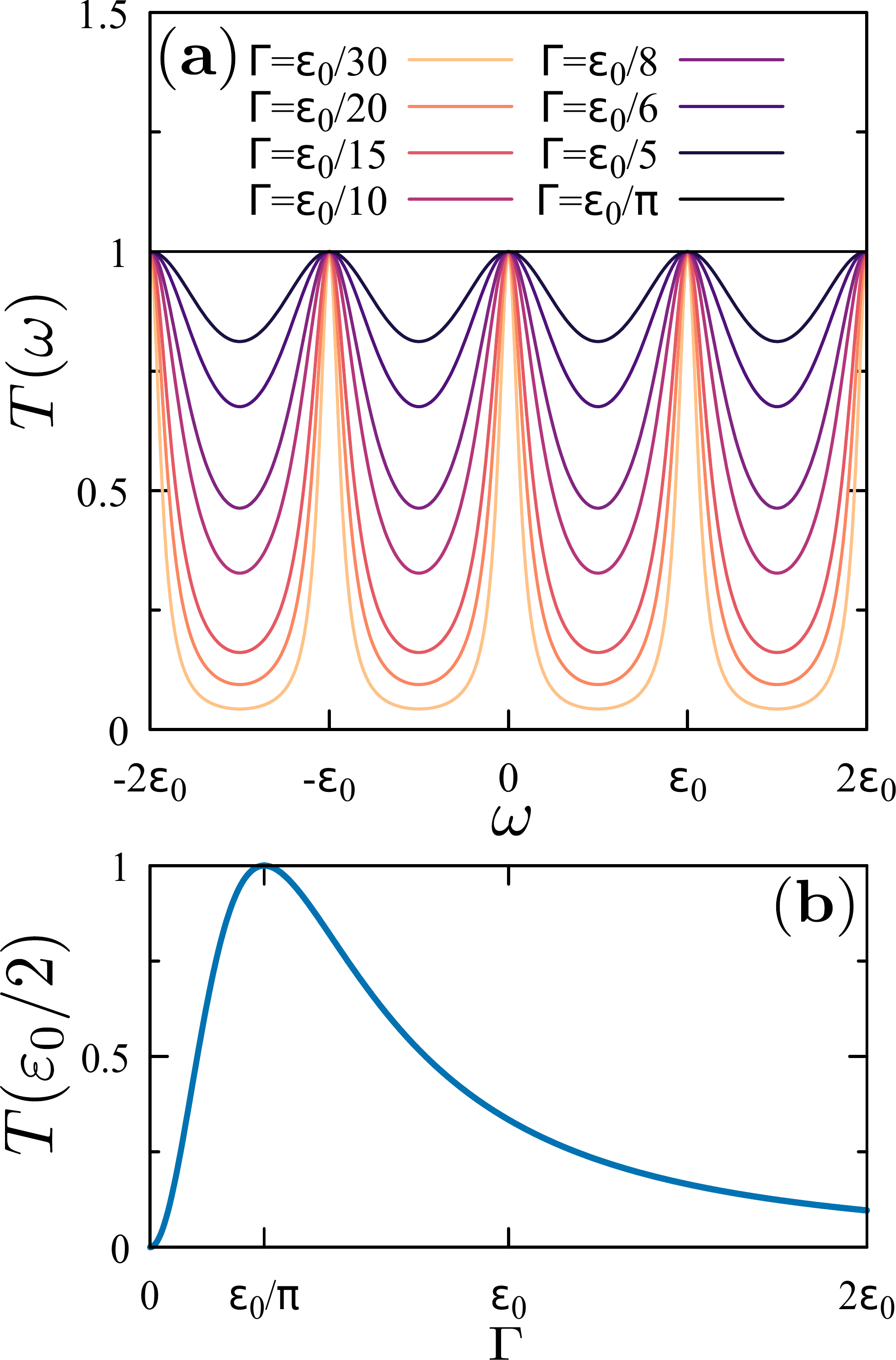}
\caption{$\bf{(a)}$ Normal transmission of the junction as a function of energy. The flux is chosen to be an integer number of flux quanta $\Phi/\Phi_0\in \mathbb{Z}$. Each curve has a different value of $\Gamma=\pi\rho(\epsilon_F)\gamma^2/\Lambda$. $\bf{(b)}$ Transmission as a function of $\Gamma$ for $\omega=\varepsilon_0/2$ and the same flux quanta as in $\bf{(a)}$.}
\label{normalT}
\end{figure}

It is worth pointing out that this transmission is completely independent of the distance between the contacts, an expected result for chiral transport. Note that there is perfect transmission $T(\omega_n)=1$ for all energies matching the eigenenergies of the uncoupled QH state $\omega_n = \varepsilon_0(n-\Phi/\Phi_0)$ with $n\in\mathbb{Z}$. There is also a particular value of $\Gamma = \varepsilon_0/\pi$ ($\tilde{\varepsilon}_0=1$) where the transmission becomes perfect for all energies and fluxes. It can be shown that this value corresponds to the one that produces a perfect matching between the chiral edge state and each lead. This condition can be thought of as a $Z=0$ barrier strength in a Blonder-Tinkham-Klapwijk (BTK) model of the junction
\citep{Blonder1982}, and will then translate into a condition of perfect Andreev reflection when the leads are considered as superconducting. In Fig.~\ref{normalT} we show a plot of the transmission as a function of energy for different hybridizations and $\Phi/\Phi_0\in\mathbb{Z}$ [Fig.~\ref{normalT}$\bf{(a)}$] and its behavior as a function of $\Gamma$ for a fixed energy $\omega=\varepsilon_0/2$ [Fig.~\ref{normalT}$\bf{(b)}$].  It is clearly seen that the transmission bears a maximum at $\Gamma=\varepsilon_0/\pi$ and the tunneling regime is recovered for $\Gamma \ll \varepsilon_0/\pi$ or $\Gamma \gg \varepsilon_0/\pi$. 
\subsection{Floquet Keldysh-Green's function method}
We here discuss the methods for the calculation of the out of equilibrium current in the Hall device when the leads are taken to be superconductors. From now on we will then work with the gauged transformed Hamiltonian defined by Eqs.~\eqref{Hnu_g}-\eqref{Hch_g}.
The time-dependent charge current flowing from the left lead to the Hall bar is obtained by means of the Heisenberg equation of motion as 
\begin{eqnarray}
\label{eqJ}
J(t) &=& e\langle\dot{\hat{N}}_L\rangle=i\frac{e}{\hbar}\langle[\hat{\widetilde{H}}(t),\sum_{\bm{k}\sigma}\hat{c}^{\dagger}_{\bm{k}L\sigma}\hat{c}^{}_{\bm{k}L\sigma}]\rangle\\
\notag
&=&i\frac{\gamma e}{\hbar}\sum_{\bm{k}\sigma}\left(e^{-i\frac{e V_b t}{\hbar}}\langle \hat{c}^{\dagger}_{\bm{k}L\sigma}\hat{\psi}^{}_{\sigma}(0)\rangle-e^{i\frac{e V_b t}{\hbar}}\langle \hat{\psi}^{\dagger}_{\sigma}(0)\hat{c}^{ }_{\bm{k}L\sigma}\rangle\right)\,,
\end{eqnarray}
where the averaged quantities $\langle \dots\rangle$ can be represented in terms of the non-equilibrium Keldysh Green's functions. In order to do so, it is practical to make use of the Nambu notation by introducing the spinors describing the QH edge state $\hat{\Psi}(s) = (\hat{\psi}_{\uparrow}^{}(s),\hat{\psi}_{\downarrow}^{\dagger}(s))^{T}$ and the superconducting leads $\hat{\upchi}_{\bm{k}\nu} = (\hat{c}_{\bm{k}\nu\uparrow}^{},\hat{c}_{\bm{-k}\nu\downarrow}^{\dagger})^{T}$. In this way, the current may be expressed in terms of $2\times 2$ matrices as
\begin{equation}
J(t)= -\frac{2 e}{\hbar}\mathrm{Re}\text{Tr}\left[\tau_z \check{\mathcal{V}}_{L0}^{}(t)\check{G}^{<}_{0L}(t,t)\right]\,,
\end{equation}
where $\tau_z$ is the Pauli matrix acting in particle-hole space, $\check{\mathcal{V}}_{L0}(t) = -\gamma e^{-i\frac{e V_b \tau_z t}{\hbar}} \tau_z$ and the lesser Green's function matrix elements are $[\check{G}^{<}_{0L}]^{\alpha\beta}(t,t) =i\sum_{\bm{k}>0}\langle\hat{\upchi}^{\dagger \beta}_{\bm{k}L}\hat{\Psi}^{\alpha}_{}(0)\rangle$. 

Even though time translation invariance is lost due to the voltage drive, the periodicity of the Hamiltonian in the period defined by the bias voltage $T = 2\pi\hbar/eV_b$ allows for the description of the Green's functions within the Floquet formalism ~\citep{Martinez2003}. In this case, all the Nambu two-time propagators can be expressed as
\begin{equation}
\check{G}(t,t') = \sum_{mn}\int_{0}^{\Omega}\frac{d\omega}{2\pi}e^{-i(\omega+m\Omega)t}e^{i(\omega+n\Omega)t'}\check{\bm{G}}^{mn}(\omega)\,,
\end{equation}
with $\Omega= 2\pi/T$. Here $\check{\bm{G}}^{mn}$ refers to the corresponding $2\times 2$ block of the Floquet Green's function $\bm{G}(\omega)$, which in principle has an infinite dimensional representation. Notice that $\check{\bm{G}}^{m+l,n+l}(\omega)=\check{\bm{G}}^{mn}(\omega+l\Omega)$, reflecting the invariance under translations in multiples of the frequency of the drive. This approach considerably simplifies the Dyson's equations of motion: When Fourier-transforming the time convolution products, the Floquet representation preserves an algebraic multiplicative structure~\citep{Tsuji2008}. After some analytical manipulations, we can finally obtain through this method the average dc current $J_{\text{DC}}=\frac{1}{T}\int_0^{T}J(t)\,dt$ as
\begin{equation}
\label{DC}
J_{\text{DC}}= -\frac{2e}{h}\mathrm{Re}\sum_{mn}\int_{0}^{\Omega} d\omega\, \text{Tr}\left[\tau_z \check{\bm{\mathcal{V}}}_{L0}^{nm}{\check{\bm{G}}}^{< mn}_{0L}(\omega)\right].
\end{equation}
The block components of the hopping in Floquet space are here defined as
\begin{eqnarray}
\label{floquet_hoppings}
\check{\bm{\mathcal{V}}}_{L0}^{nm} &=& \frac{1}{T}\int_{0}^{T}e^{i(n-m)\Omega t}\check{\mathcal{V}}_{L0}(t)\, dt\\
\notag
&=&\check{\mathcal{V}}_{-}\,\delta_{m,n+1}+\check{\mathcal{V}}_{+}\,\delta_{m,n-1}\,,
\end{eqnarray}
with $\check{\mathcal{V}}_{\pm} = \mp\frac{\gamma}{2}(\tau_0\pm\tau_z)$. The block elements of the lesser Green's function $\check{\bm{G}}^{< mn}_{0L}(\omega)$ can be easily obtained by applying the Langreth rules in the Floquet-Dyson equations of motion~\citep{Jauho1996}. All that is eventually needed are the uncoupled Green's functions of the chiral edge state and the superconducting leads which, within our model, have closed analytical expressions. We refer the reader to Appendix \ref{AppendixA} for technical details of the calculation of the Floquet Green's functions.
\section{Current-voltage characteristics}\label{III}
\begin{figure}[b]
\includegraphics[width=0.9\columnwidth]{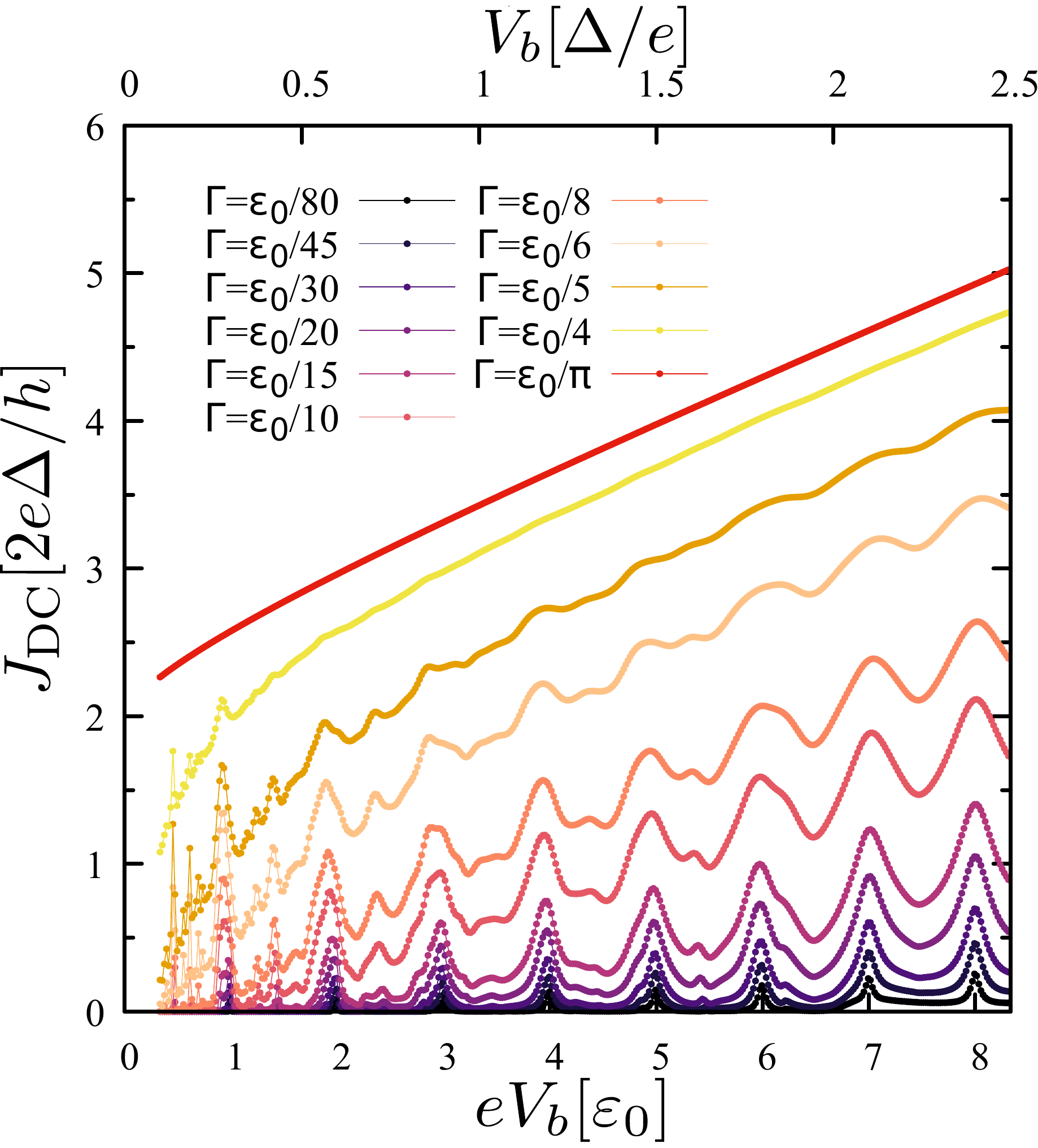}
\caption{Current-voltage characteristics for an integer number of flux quanta threading the sample $\Phi/\Phi_0 \in \mathbb{Z}$ and different hybridization parameters $\Gamma=\pi\rho(\varepsilon_F)\gamma^2/\Lambda$ expressed as a fraction of the level spacing $\varepsilon_0=0.3\,\Delta$.}
\label{fig3}
\end{figure}

\begin{figure*}[t]
\includegraphics[width=\textwidth]{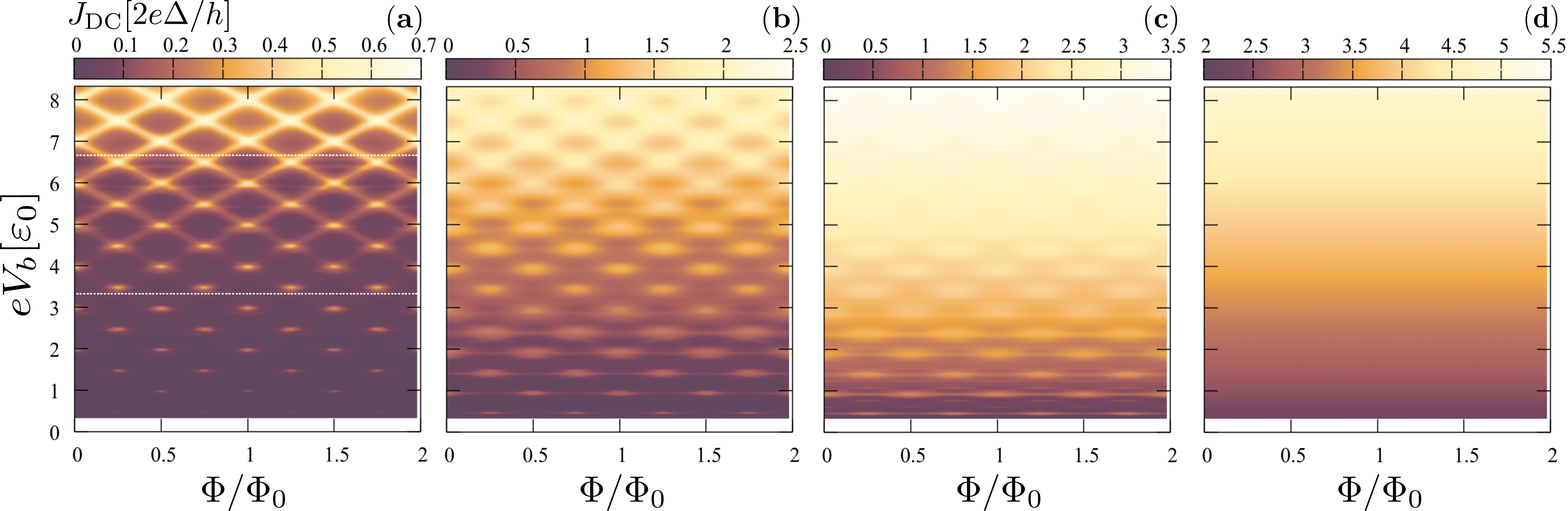}
\caption{Time averaged current $J_{\text{DC}}$ as a function of the bias voltage $V_b$ and the number of fluxes enclosed by the edge state $\Phi/\Phi_0$. Each color map has been calculated with a different hybridization $\Gamma$ of the edge modes with the leads: $\bf{(a)}$ $\Gamma=\varepsilon_0/30$, $\bf{(b)}$ $\Gamma=\varepsilon_0/10$, $\bf{(c)}$ $\Gamma=\varepsilon_0/6$ and $\bf{(d)}$ $\Gamma=\varepsilon_0/\pi$. In $\bf{(a)}$, we highlight with white dashed lines the bias voltages corresponding to $eV_b=\Delta$ and $eV_b=2\Delta$.}
\label{fig4}
\end{figure*}
We here discuss the main results of our work, namely, the transport simulations of the voltage biased Josephson junction. From here on we will focus on the zero-temperature limit. In Fig.~\ref{fig3} we show the time-averaged dc current as obtained by numerically evaluating Eq.~\eqref{DC} when varying the bias voltage $V_b$ in the case where there is an integer number of fluxes threading the sample $\Phi/\Phi_0 \in \mathbb{Z}$.  Each of the curves has a different hybridization $\Gamma = \pi\rho(\epsilon_F)\gamma^2/\Lambda$ expressed as a fraction of the level spacing $\varepsilon_0$  [see Eq.~\eqref{Hch}], which is here fixed at $\varepsilon_0 = 0.3\,\Delta$. 
This choice of parameters is consistent with the typical orders of magnitude in experiments done with graphene and MoRe contacts~\citep{Amet2016,Draelos2018}, where $v_d\sim 10^6\,\text{m/s}$, $\Lambda\sim 10\,\mu\text{m}$ and the pairing gap $\Delta\simeq 1.3\,\text{meV}$.
In the lower $x$ axis the voltage is normalized to $\varepsilon_0$, and in the upper axis it is normalized to the superconducting gap $\Delta$, so as to properly compare both scales. We have checked that the current-voltage characteristic is completely independent of the angle $\alpha$ appearing in the nonlocal propagators. This is a consequence of the coherent and chiral quasiparticle transport, which makes the differential conductance of the junction independent of the distance between the superconducting contacts. As a matter of fact, it can be shown that the distance between leads plays the role of a superconducting phase difference $\varphi_0 = 2(\pi-\alpha)\Phi/\Phi_0$, which fixes a time origin in the two-terminal out of equilibrium setup and hence becomes irrelevant for the time averaged dc current. 

For low hybridizations a rich subgap structure is apparent from the Fig.~\ref{fig3}, where a series of resonances occur each time the bias $eV_b$ becomes a multiple (or harmonic) of the level spacing $\varepsilon_0$. As $\Gamma$ grows larger, these features wash out until the current-voltage characteristic reaches a completely transparent limit for $\Gamma =\varepsilon_0/\pi$. For hybridizations such that $\Gamma\gg \varepsilon_0/\pi$ or $\Gamma \ll \varepsilon_0/\pi$ the tunneling regime is recovered, in accordance with the behavior of the normal transmission of the junction (see Fig.
~\ref{normalT}). Although not shown, we have also verified that in the limit $\varepsilon_0\gg \Delta$ we obtain the well known transport results of voltage-biased junctions with single-level quantum dots~\citep{Averin1995,Yeyati1997,Johansson1999}. In the opposite limit, $\varepsilon_0\ll\Delta$, the dc current becomes rather featureless as a function of voltage but preserves a periodic structure with the superconducting flux quantum.

The results shown in Fig.~\ref{fig4} are quite different from the well-known current-voltage characteristic of other Josephson junctions, where resonant features are expected to be present at bias voltages which are subharmonics of the pairing gap $e V_b=2\Delta/n$~\citep{Cuevas1996,Averin1995,Meyer2011,SanJose2013}. In our case, the subgap structure provides more information on the discreteness of the chiral  edge modes than on the BCS singularity. These findings may be compared with the ones obtained in a recent work~\citep{Bezuglyi2017} where a detailed analysis of the current-voltage characteristic of a long junction was presented, though in a quite different setup. When the junction's length
greatly exceeds the coherence length $\xi$ of the superconductor---typically of the order of a few to a few hundred nanometers---it has been found that the current shows characteristic  peaks each time the bias voltage is a multiple of the distance between the static Andreev levels. This is expected to be precisely the case in a QH based junction, where the effective length is given by the perimeter of the Hall bar $\Lambda$, which is usually a few micrometers and hence $\Lambda\gg \xi$. 
\begin{figure}[b]
\includegraphics[width=0.85\columnwidth]{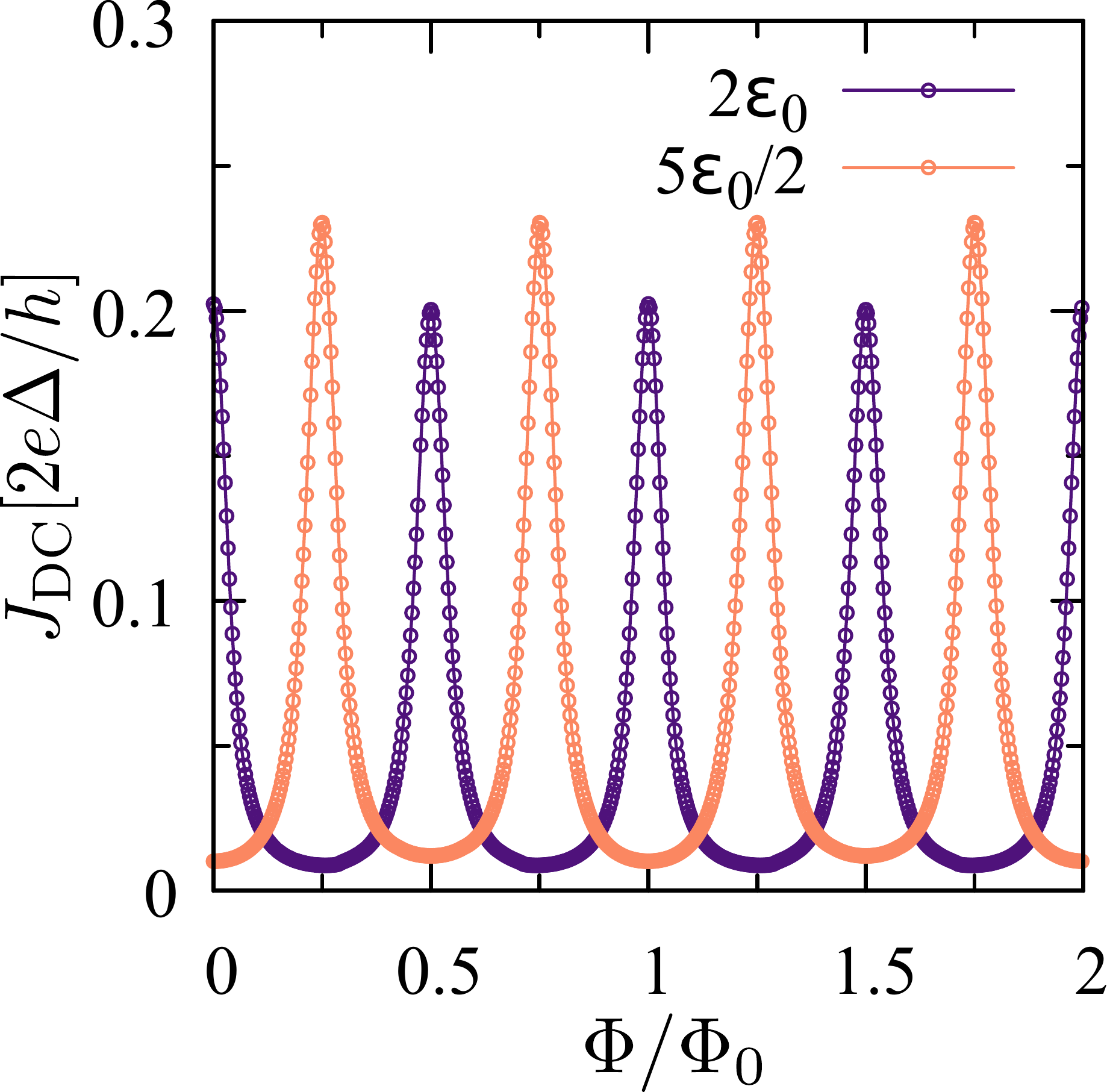}
\caption{Time-averaged dc current at bias voltages $e V_b = 2\varepsilon_0$ and $e V_b = 5\varepsilon_0/2$ as a function of the flux enclosed by the edge state. The hybridization is $\Gamma=\varepsilon_0/30$, as in Fig.~\ref{fig4}$\bf{(a)}$. Note that the periodicity is determined by the superconducting flux quantum $\Phi_0^{s}=\Phi_{0}/2$.}
\label{fig5}
\end{figure}
 
\begin{figure*}[t]
\includegraphics[width=\textwidth]{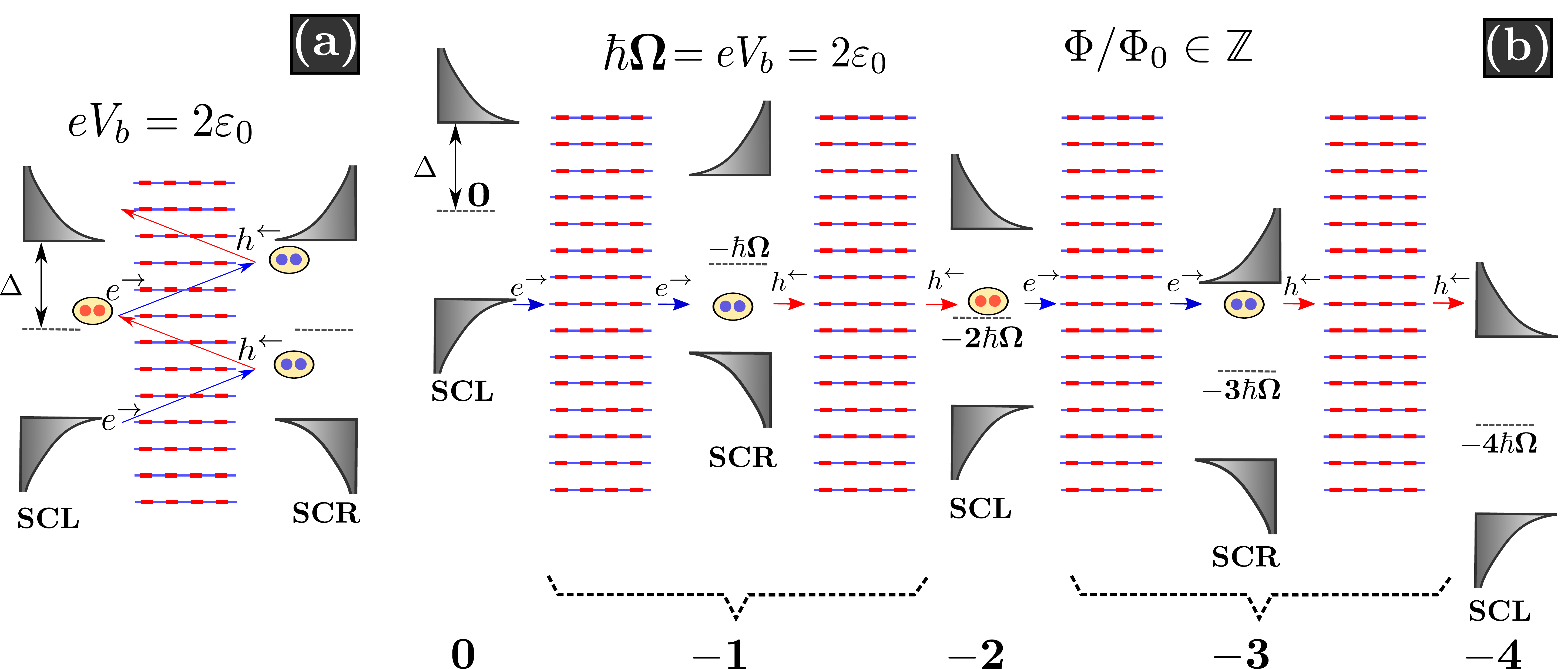}
\caption{$\mathbf{(a)}$ Usual pictorial scheme of the multiple Andreev reflection processes that contribute to the quasiparticle current. The bias voltage is such that $eV_b=2\varepsilon_0$, and the flux enclosed by the edge state is an integer number of flux quanta. Blue arrows correspond to right-moving electronic states, and red arrows correspond to left-moving holelike states. Short solid (dashed) lines indicate the position of the uncoupled electron (hole) states. The path is to be understood sequentially. Each time an Andreev process occurs, a Cooper pair is transferred to the corresponding condensate. $\mathbf{(b)}$ Alternative diagram of the MAR trajectories in the Floquet space of replicas. This is merely an unfolding of $\mathbf{(a)}$, so that the interpretation of the transport process can be regarded at a fixed energy. Each time a right-moving electron or a left-moving hole is transferred from one lead to the other, it switches to a different replica of the superconducting lead. The replica index of the QH region and each superconductor is specified in the lower part of the figure. SCL and SCR refer to the left and right superconducting leads, respectively.}
\label{ladders}
\end{figure*}
The fact that the current is carried by QH edge channels offers an additional handle: the chirality imposed by the magnetic flux. To better visualize how these resonances evolve when varying the flux enclosed by the edge state, we show in Fig.~\ref{fig4} color maps of the dc current as a function of both the bias voltage $V_b$ and the number of flux quanta $\Phi/\Phi_0$ threading the sample (recall that in our model, $\Phi$ actually represents a departure from a reference flux). Each panel has been calculated for a given value of $\Gamma$. For small couplings, such as the one shown in Fig.~\ref{fig4}$\bf{(a)}$, it is possible to appreciate a series of well defined peaks at low voltages that are located at
\begin{equation}
\label{resoVb}
  e V_b=\begin{cases}
    n\,\varepsilon_0 & \text{if $\Phi/\Phi_0 \in \mathbb{Z}$ or $\mathbb{Z}+\frac{1}{2}$}\\
    (n+\frac{1}{2})\,\varepsilon_0, & \text{if $\Phi/\Phi_0 \in \mathbb{Z} \pm \frac{1}{4}$},
  \end{cases}
\end{equation}
with $n$ being an integer number. When the bias voltage $e V_b > \Delta$, these resonances tend to disperse linearly with the flux variations along the curves defined by 
\begin{equation}
\label{resoCooper}
e V_b = \varepsilon_0(n \pm 2\Phi/\Phi_0)\,.
\end{equation}

These well-defined peaks are blurred as the hybridization grows larger [see panels Fig.~\ref{fig4}$\bf{(b)}$ and Fig.~\ref{fig4}$\bf{(c)}$], in accordance with the general tendency observed in Fig.~\ref{fig3}. When the transparent limit is reached, the current-voltage characteristic becomes completely independent of the flux variations in the sample, as shown in Fig.~\ref{fig4}$\bf{(d)}$. For this particular hybridization, the flux accumulated by an electron is completely canceled out by the perfectly Andreev-reflected hole resulting in a flux-independent current.

We show in Fig.~\ref{fig5} two horizontal cuts of Fig.~\ref{fig4}$\bf{(a)}$
when the bias voltage is tuned at $e V_b = 2\varepsilon_0$ and $e V_b = 5\varepsilon_0/2$. This plot captures not only the resonant behavior but also the clear periodicity of the quasiparticle current with the superconducting flux quantum $\Phi_0^{s}=\Phi_0/2$, which is in stark contrast to the $\Phi_0$ periodicity of the equilibrium non-dissipative Josephson supercurrent \cite{Ma1993,vanOstaay2011,Stone2011}.

To understand the appearance of this structure, it is useful to regard the elementary tunnel processes  (for small hybridizations) which give rise to the quasiparticle current in the junction. To this end, we show in Fig.~\ref{ladders} two different schemes to analyze the existence of resonant trajectories that eventually lead to the peaks in the current. We illustrate the case of a bias voltage $e V_b=2\varepsilon_0$ and an integer number of flux quanta $\Phi/\Phi_0 \in \mathbb{Z}$. The short solid and dashed lines represent the uncoupled electronic ($-$) and hole ($+$) states, respectively, arising from the QH edge which are degenerate for these particular values of the parameters. This set of discrete levels is well described by the spectrum of Eq.~\eqref{Hch_g}, given by $E_n^{\mp}(\Phi)=\varepsilon_0(n\mp\Phi/\Phi_0)\mp e V_b/2$. The static Andreev bound states living inside the BCS gap are essentially located at these energies for zero bias and very small hybridizations ($\Gamma\ll \varepsilon_0$). 
In Fig.~\ref{ladders}$\mathbf{(a)}$ we show the usual diagram of the multiple Andreev reflection processes taking place throughout the gapped region: Right-moving electrons ($e^{\rightarrow}$) and left-moving holes ($h^{\leftarrow}$) gain an energy $eV_b$ when flowing from one terminal to the other until they reach the continuum spectrum. Note that this sketch is consistent with the bias voltage being gauged away from the leads to the time-dependent tunneling elements. For this choice of parameters, there is a perfect ladder of equally spaced states in the QH region, leading to quasiparticle transfer through trajectories which always cross a resonant level. In junctions with single-level quantum dots, it has been argued that this condition results in an enhancement of the dc current~\citep{Yeyati1997}.

An alternative description of the same process is shown in Fig.~\ref{ladders}$\mathbf{(b)}$, which is basically an unfolding of the MAR trajectories shown in  Fig.~\ref{ladders}$\mathbf{(a)}$. In this schematic, we take advantage of the Floquet space to represent the transport of quasiparticles between superconductors as a fixed-energy process. Each time an electron or a hole is transferred from one terminal to the other, it switches to a consecutive replica of the BCS superconducting leads (see Fig.~\ref{fig_app} in Appendix \ref{AppendixA} for details). Within this scheme, it is clearly seen that the resonant trajectories bear a suggestive resemblance to a resonant transfer in a multibarrier structure, in a similar fashion to the mapping discussed in Ref.~\citep{Johansson1999-bis}. Indeed, when the bias and fluxes are precisely tuned, the electronic and hole states can be resonantly transmitted via the edge modes of the chiral state throughout the whole process. When analyzing the transport of electrons from the left to the right lead, a series of conditions must be satisfied for the resonance to take place.  It is necessary to have degenerate electron and hole states at the $l$th replica of the Hall device. Additionally, there must be an electronic state in the $l-2$ replica with the same energy as that of the holelike state in the $l$th replica, which guarantees the resonant condition throughout the entire path, as can be seen from Fig.~\ref{ladders}$\mathbf{(b)}$. This set of equations reads
\begin{eqnarray}
\label{cond1}
E^{-}_{n,l}(\Phi)&=&E^{+}_{m,l}(\Phi)\\
E^{+}_{m,l}(\Phi)&=&E^{-}_{r,l-2}(\Phi)\,,
\label{cond2}
\end{eqnarray}
where we define the spectrum of the $l$th Floquet  replica as $E^{\mp}_{n,l}(\Phi)=\varepsilon_0(n\mp \Phi/\Phi_0)\mp eV_b/2+leV_{b}$. The fluxes and voltages which satisfy these constraints are precisely given by Eq. ~\eqref{resoVb}. When taking into account the mirror processes of transferring holes from the right to the left lead, the same result can be found, indicating the degeneracy between these two mechanisms. When these conditions are fulfilled, there is a natural enhancement of the charge-transfer process which generates, in turn, a well-defined peak in the dc current even at very low voltages. As the hybridization grows larger, the decoupled picture breaks down leading to a broadening of these Lorentzian-shaped resonances. This mechanism of resonant transfer may be compared to what occurs in junctions involving single
~\cite{Yeyati1997} or several discrete levels~\citep{Dolcini2008}. The main difference is that the linear dispersion of the edge state inherently provides a discrete set of states which may be tuned to be completely equidistant from each other, allowing for the resonant ladder to take place~\cite{Bezuglyi2017}. Alternatively, it can be thought of as a perfect alignment of all resonant modes in the Floquet multibarrier picture. It is also worth noticing that in this QH setup it is not necessary to employ gate voltages in the normal region to manipulate the spectrum, since the flux enclosed by the chiral edge state controls the relative position of the electronic and holelike levels.

For $e V_b>\Delta$, there is a direct process to transfer Cooper pairs. This process is related to the so-called resonant Cooper pair transfer mechanism that occurs in junctions with single-level quantum dots when the chemical potential of either of the leads is aligned with the bound state~\citep{Johansson1999,Buitelaar2003}. Under these circumstances, an electronic state can tunnel resonantly through this level and the Andreev-reflected hole as well, effectively transferring a Cooper pair from one lead to the other.
In our case, for this two-step trajectory to become resonant, it is only necessary to ask for a unique constraint between the voltage and the flux enclosed by the edge channel, leading to peaks in the dc current that disperse linearly with flux [cf. Eq.~\eqref{resoCooper}]. In fact, when taking into account the processes that tunnel electrons or holes, one finds that Eqs.~\eqref{cond1} and \eqref{cond2} independently lead to a current enhancement in agreement with the condition  given by Eq.~\eqref{resoCooper}. 

Interestingly, a full reconstruction of the dc current in the flux-voltage plane allows for a complete spectroscopy of the chiral modes. Indeed, for $eV_b>\Delta$ the resonant Cooper pair transfer processes provide a clear trail of the dispersion of the edge state with flux, allowing for a direct characterization of its drift velocity.
\begin{figure}[t]
\includegraphics[width=0.9\columnwidth]{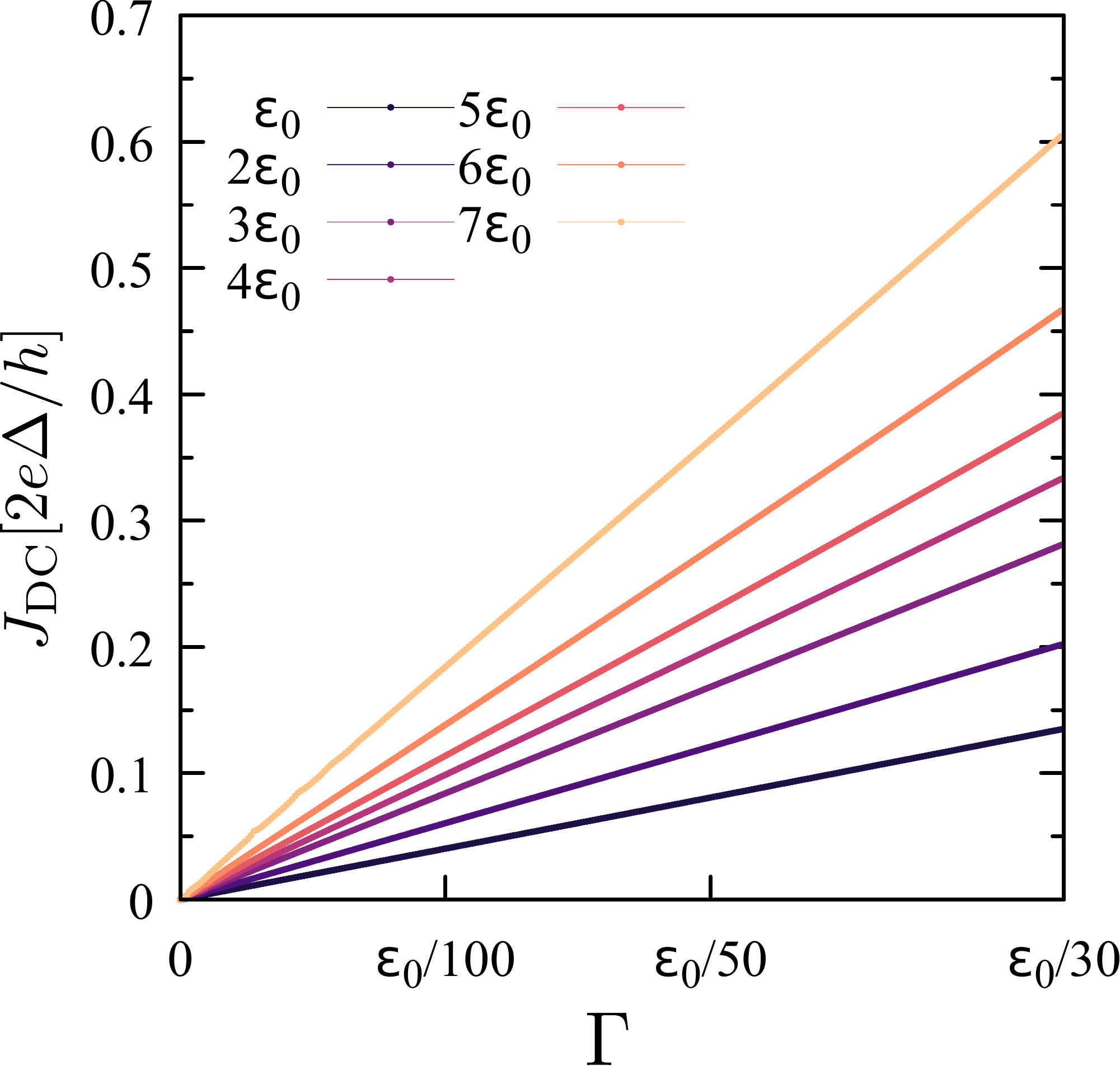}
\caption{Dependence of the quasiparticle current peaks for low hybridizations as a function of $\Gamma$ for an integer number of fluxes threading the system. Each curve has been calculated for different values of the resonant bias voltage $eV_b=n\varepsilon_0$, and thus represents a different peak.}
\label{mars_gamma}
\end{figure}

\begin{figure*}[t]
\includegraphics[width=\textwidth]{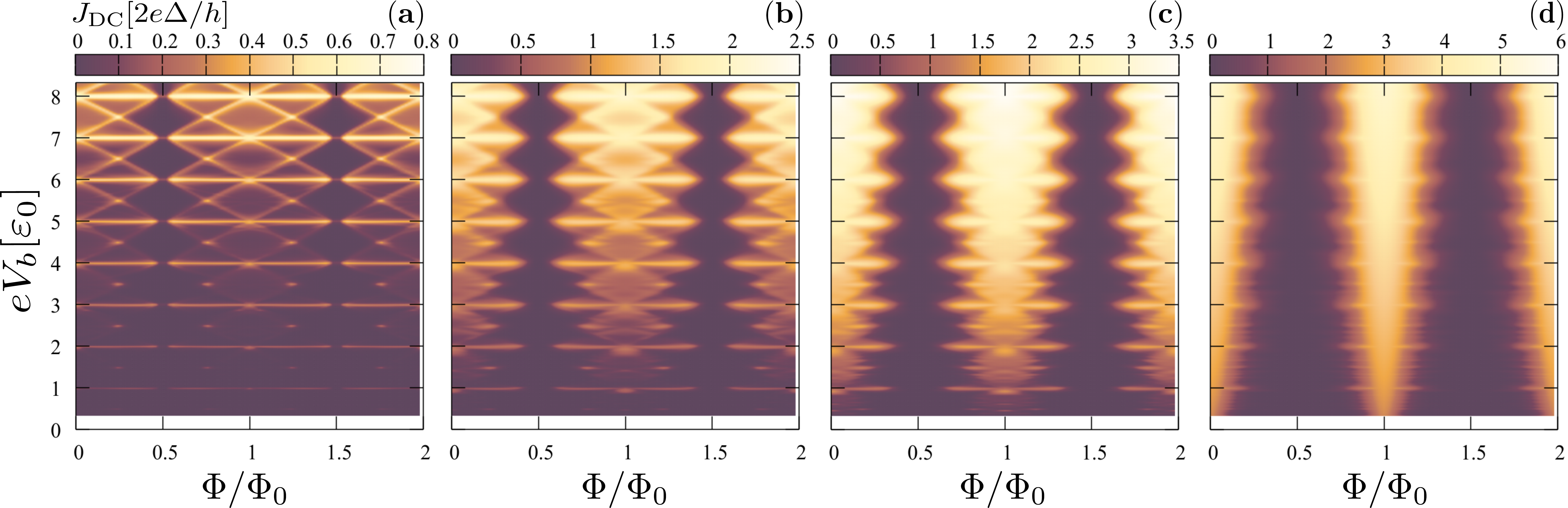}
\caption{Time averaged current $J_{\text{DC}}$ as a function of the bias voltage $V_b$ and the number of fluxes enclosed by movers with left chirality $\Phi/\Phi_0$ in the Aharonov-Bohm setup where the central Hamiltonian is described by Eq.~\eqref{HAB}. The movers with right chirality enclose the opposite flux. Each color map has been calculated with a different hybridization $\Gamma$ of the edge modes of each chirality with the leads: $\bf{(a)}$ $\Gamma=\varepsilon_0/60$, $\bf{(b)}$ $\Gamma=\varepsilon_0/20$, $\bf{(c)}$ $\Gamma=\varepsilon_0/12$ and $\bf{(d)}$ $\Gamma=\varepsilon_0/2\pi$.}
\label{fig_nonchiral}
\end{figure*}
Having identified the resonant processes that lead to the quasiparticle current enhancement, it is instructive to look at the scaling of each peak with the hybridization parameter $\Gamma$. In Fig.~\ref{mars_gamma} we show the evolution of the height of the dc current peaks as a function of $\Gamma$ for an integer number of fluxes threading the sample. Each curve has been calculated for different values of the resonant voltage $eV_b = n\varepsilon_0$ and hence represents the magnitude of the $n$th current peak ($n=1, ..., 7$) shown in Fig.~\ref{fig3}, which is well defined for $\Gamma\ll \varepsilon_0/\pi$. Notably, every resonance scales linearly with $\Gamma$, a fact which stands in contrast to the usual picture of quasiparticle transport between superconductors in the tunneling regime. Indeed, in Josephson junctions with weak links, $\mathrm{N}$ quasiparticles are expected to be shuttled from one lead to the other in the voltage bias range defined by $2\Delta/\mathrm{N} < eV_b < 2\Delta/\mathrm{N}-1$, giving rise to a dc current with a dominant order scaling given by $\Gamma^{\mathrm{N}}$~\citep{Cuevas1996}. The linear scaling in $\Gamma$ is known to occur in junctions with single-level quantum dots only when Cooper pairs are transferred resonantly through the single level. In fact, by analyzing the equivalent three-barrier structure in energy space, Ref.~\citep{Johansson1999} reported that the height of these peaks in the dc current is proportional to the hybridization. In our case, the universal lineal scaling of \textit{all} current peaks is a peculiarity of the resonant transfer of quasiparticles through out the whole Floquet multibarrier structure. Indeed, all the energy-conserving processes, like the one depicted in Fig.~\ref{ladders}$\mathbf{(b)}$, lead to the same scaling independently of the number of barriers in the MAR path. This is another distinguishing feature of the linear spectrum of the chiral edge state, which provides a discrete set of equidistant energy levels which may be tuned with magnetic flux to allow for the resonant (order $\Gamma$) quasiparticle transport to occur.

\section{Comparison with non-chiral transport}\label{IV}
A natural follow-up question is whether these results would change if the system were to admit the existence of counter-propagating edge states. In this section, we will briefly discuss the main differences in the current-voltage characteristics when both modes of left chirality ($\hat{\psi}_{n_L\sigma}$) and right chirality ($\hat{\psi}_{n_R\sigma}$)  are present in the system. To this end, we redefine the central region to be described with a Aharonov-Bohm-like Hamiltonian given by
\begin{eqnarray}
\notag
\hat{H}_{\text{A-B}}&=&\sum_{n_L \sigma }\varepsilon_0\left(n_L-\frac{\Phi}{\Phi_0}\right)\hat{\psi}^{\dagger}_{n_L\sigma}\hat{\psi}_{n_L\sigma}^{}\\
&&+\sum_{n_R\sigma}\varepsilon_0\left(n_R+\frac{\Phi}{\Phi_0}\right)\hat{\psi}^{\dagger}_{n_R\sigma}\hat{\psi}_{n_R\sigma}^{}\,.
\label{HAB}
\end{eqnarray}
It is then straightforward to generalize the electronic local and non-local retarded Green's functions to this case as
\begin{equation}
g_{00}^{r}=\frac{\pi}{\Lambda \varepsilon_0}\left[\cot\left(\pi\frac{\omega+i\eta}{\varepsilon_0}+\pi\frac{\Phi}{\Phi_0}\right)+\cot\left(\pi\frac{\omega+i\eta}{\varepsilon_0}-\pi\frac{\Phi}{\Phi_0}\right)\right]\,,
\label{g00_nc}
\end{equation}
with $g^{r}_{\alpha\alpha}=g_{00}^{r}$ and
\begin{eqnarray}
\notag
g_{0\alpha}^{r}&=&\frac{\pi}{\Lambda\varepsilon_0}\bigg[e^{i(\pi-\alpha)\left(\frac{\omega+i\eta}{\varepsilon_0}+\frac{\Phi}{\Phi_0}\right)}\csc\left(\pi\frac{\omega+i\eta}{\varepsilon_0}+\pi\frac{\Phi}{\Phi_0}\right)\\
&+&e^{-i(\pi-\alpha)\left(\frac{\omega+i\eta}{\varepsilon_0}-\frac{\Phi}{\Phi_0}\right)}\csc\left(\pi\frac{\omega+i\eta}{\varepsilon_0}-\pi\frac{\Phi}{\Phi_0}\right)\bigg].
\label{g0alpha_nc}
\end{eqnarray}
The propagator $g_{\alpha 0}^{r}$ is obtained by changing $\alpha\rightarrow 2\pi - \alpha$ in $g_{0\alpha}^{r}$. The normal transmission of this model can be obtained by simply replacing the uncoupled propagators defined by Eqs.~\eqref{g00_nc} and \eqref{g0alpha_nc} in Eq.~\eqref{T}. In this model, the transmission is strongly dependent on the angle $\alpha$ that determines the distance between the leads, as opposed to the chiral model. For the sake of concreteness, we shall work with the most symmetric case, namely, $\alpha=\pi$. In such circumstances, the normal transmission is exactly zero for half an integer number of fluxes threading the system ($\Phi/\Phi_0 \in \mathbb{Z} + 1/2$), since the non-local propagator vanishes exactly [see Eq. \eqref{g0alpha_nc}]. 

When considering the driven system, we will make use of the same gauge transformation as the one described in Sec. \ref{IIA} so that the central Hamiltonian changes to
\begin{equation}
\widetilde{H}_{\text{A-B}}\!=\!H_{\text{A-B}} - \frac{eV_b}{2}\left(\sum_{n_R,\sigma}\hat{\psi}^{\dagger}_{n_R\sigma}\hat{\psi}_{n_R\sigma}^{}\!+\!\sum_{n_L,\sigma}\hat{\psi}^{\dagger}_{n_L\sigma}\hat{\psi}_{n_L\sigma}^{}\right)\,,
\end{equation}
and the time dependence is solely included in the hopping to the left superconductor. The Floquet spectrum of the uncoupled Aharonov-Bohm ring is then defined by electronic and hole states with different chirality as $E_{n_L,l}^{\mp}=\varepsilon_0(n_L \mp \Phi/\Phi_0)\mp eV_b/2 + leV_b$ and $E_{n_R,l}^{\mp}=\varepsilon_0(n_R \pm \Phi/\Phi_0)\mp eV_b/2 + leV_b$, with $l$ being the replica index.

\begin{figure}[t]
\includegraphics[width=0.8\columnwidth]{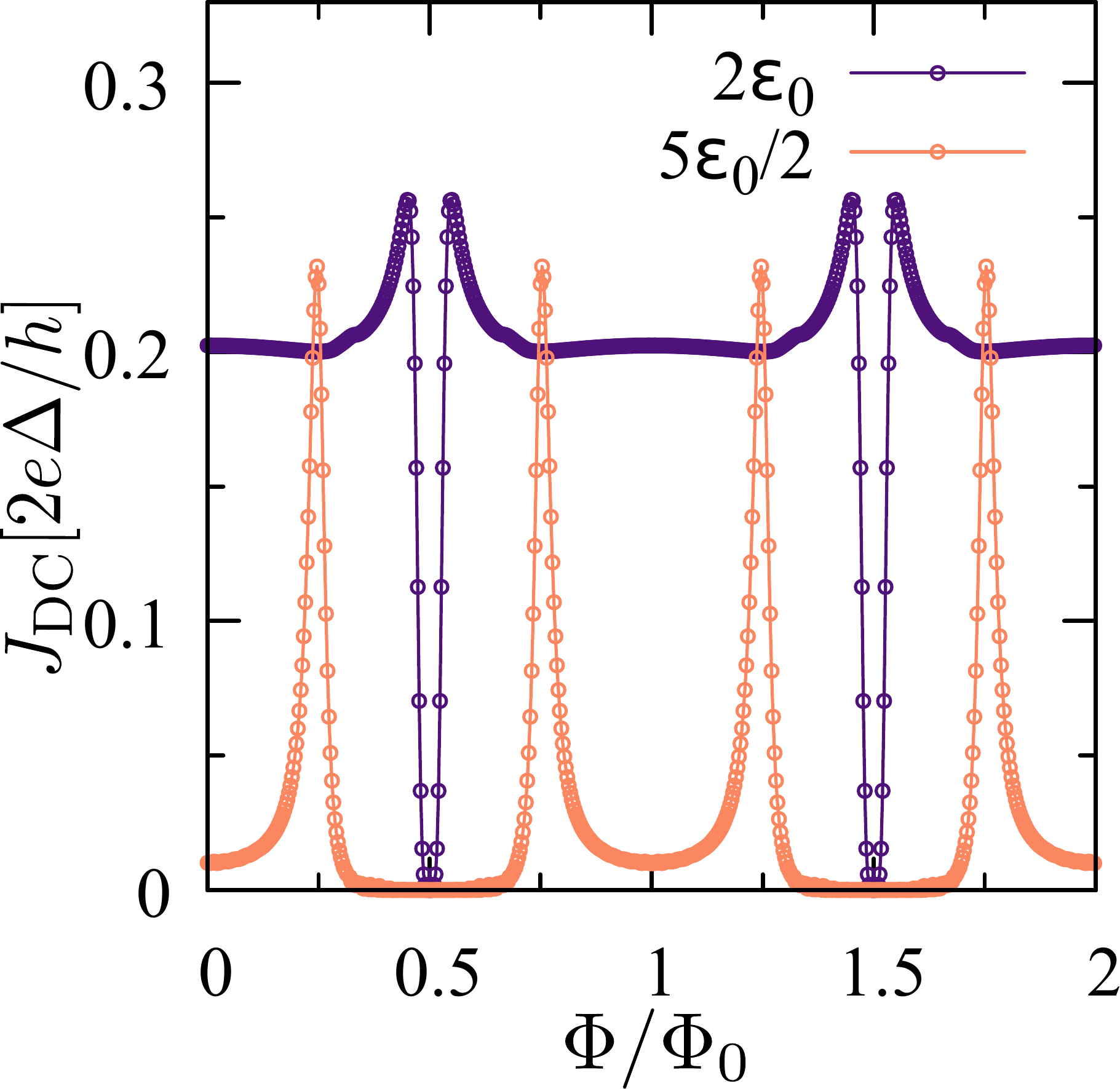}
\caption{Time averaged dc current of the non chiral model at bias voltages $e V_b = 2\varepsilon_0$ and $e V_b = 5\varepsilon_0/2$ as a function of the flux enclosed by movers with left chirality. The hybridization is $\Gamma=\varepsilon_0/60$, as in Fig.~\ref{fig_nonchiral}$\bf{(a)}$.}
\label{fig9}
\end{figure}

We show in Fig.~\ref{fig_nonchiral} the current-voltage maps of this non-chiral model as a function of the flux enclosed by the movers with left chirality. Each panel has been calculated for a different hybridization of the edge modes of each chirality with the superconducting terminals. Since the number of channels is increased by a factor of 2 with respect to the chiral setup, we show color maps with hybridization parameters which are half the ones used in Fig.~\ref{fig4} to better compare both results. A first clear difference with the chiral model is the suppression of the dc current whenever $\Phi/\Phi_0\in \mathbb{Z}+1/2$ due to the existence of destructive interfering paths in the device. On the other hand, the presence of counter propagating states gives rise to new resonances, which are mainly the horizontal lines in Fig.~\ref{fig_nonchiral} at bias voltages which are multiples of the discrete level spacing. When considering the system parameters that allow for a complete alignment of the resonant levels in the Floquet multibarrier picture, a constraint similar to the one described for the chiral model [Eqs.~\eqref{cond1} and ~\eqref{cond2}] is found when taking into account separately the movers of each chirality. The new condition emerges from backscattering processes where the electronic channels with left (right) chirality become degenerate with hole-like channels with right (left) chirality in the same replica, which automatically ensures the resonance throughout the whole MAR path. In this case the alignment condition reads
\begin{equation}
E^{-}_{n_L,l}(\Phi)=E^{+}_{n_R,l}(\Phi),    
\end{equation}
which is satisfied for $eV_b = n\varepsilon_0$ with $n\in \mathbb{Z}$ and thus explains the extra horizontal resonances. We show in Fig.~\ref{fig9} two horizontal cuts of Fig.~\ref{fig_nonchiral}$\mathbf{(a)}$ when the bias voltages are tuned to be $eV_b=2\varepsilon_0$ and $eV_b=5\varepsilon_0/2$. The exact suppression of the dc current at half an integer number of fluxes can be appreciated for both curves.  On the other hand, for $eV_b=2\varepsilon_0$ the resonant features are dominated by backscattering in the device bearing significant deviations from the quasiparticle current in the chiral model at the same bias voltage (see Fig.~\ref{fig5}). When $eV_b=5\varepsilon_0/2$, the chiral paths are the ones that dictate the appearance of resonant peaks, and hence the current has a resemblance to that of Fig.~\ref{fig5}, setting aside the suppression near the half-integer quanta. 

It is then clear that it is possible to distinguish whether the dissipative transport of quasiparticles is mediated by chiral or non-chiral edge channels by performing current-voltage measurements as a function of the flux threading the sample. The main differences are due to the existence of backscattering channels which are responsible for both the suppression of the dc current at particular values of the fluxes and for the emergence of new resonant paths.
\section{Summary and conclusions}\label{V}
We have presented numerical calculations of the current-voltage characteristic of a SC-QH-SC junction where the dissipative current is carried by a single spin-degenerate chiral edge channel. The existence of resonant MAR trajectories produces a  distinctive subgap structure in the dc current which is periodic with the superconducting flux quantum $\Phi_0^{s}=\Phi_0/2$, as opposed to the non-dissipative Josephson supercurrent which is periodic with the normal flux quantum $\Phi_0$.

The presence of a chiral edge state with a constant drift velocity ensures the existence of an equidistant spectrum of electronic and hole modes in the normal region with a relative position which may be tuned with magnetic flux to a condition where quasiparticles are resonantly transferred through the junction. By means of an interpretation of the transport processes as occurring in a multibarrier structure in Floquet space, we have identified the specific values of bias voltages where an enhancement of the dc current should be expected as a function of the flux variations in the device. Under these circumstances, the well defined peaks developed in the current-voltage characteristic scale linearly with the hybridization parameter $\Gamma$, a manifestation of the resonant transport of quasiparticles between superconductors. For voltages $e V_b>\Delta$, the resonant Cooper pair tunneling processes give rise to peaks that disperse with flux providing information on the drift velocity of the quantum Hall edge channel. We expect our results to remain valid for finite temperatures as long as $k_B T \ll \varepsilon_0$, so that thermalization effects inside the Hall sample can be safely ignored---this guarantees that the coherence length is larger than the sample’s perimeter~\citep{Zhao2020}.

Throughout this paper we have neglected the Zeeman splitting between the two occupied spin species of the chiral channel. Nonetheless, one can easily check that, since the electronic levels of a given spin are shifted in the same amount as the holelike levels of the opposite spin, the addition of a Zeeman term does not alter the resonant condition determined by Eq.~\eqref{cond2}. Although not shown, we have numerically verified that the position of the main peaks in the current-voltage characteristic remains unaltered when taking into account this spin-splitting term.

We have furthermore analyzed the role of chirality by comparing these transport results with an Aharonov-Bohm setup where backscattering between movers of opposite chirality is allowed. We found clear differences in the current-voltage characteristics as a function of magnetic flux, such as the presence of destructive interference paths that suppress the current at certain fluxes and the appearance of additional resonant MAR trajectories which lead to new peaks in the dc current.

These results may be considered as a first step towards the understanding of nonequilibrium transport in Josephson junctions bridged by quantum Hall edge channels. A complete tomography of the dissipative current as a function of the flux threading the sample and the bias voltage between superconductors could also be used as evidence of chiral mediated transport in these hybrid devices.
\begin{acknowledgments}
We  acknowledge financial support from ANPCyT (Grants No. PICTs 2016-0791 and No. 2018-01509), from CONICET (Grant No. PIP 11220150100506) and from SeCyT-UNCuyo (Grant No. 2019 06/C603).
\end{acknowledgments}

\appendix
\section{Green's functions in Floquet representation}\label{AppendixA}
We present in this appendix a detailed discussion of the Floquet-Green's function method for the calculation of the lesser Green's functions between the leads and the chiral state, which are ultimately needed to obtain the current flowing in the device. As stated in the main text, these may be obtained by applying the Langreth rules in the Floquet-Dyson equation of motion, as we will show in the following. All that is eventually needed are the uncoupled Green's functions of the edge state and the superconducting terminals.

\begin{figure}[t]
\includegraphics[width=0.95\columnwidth]{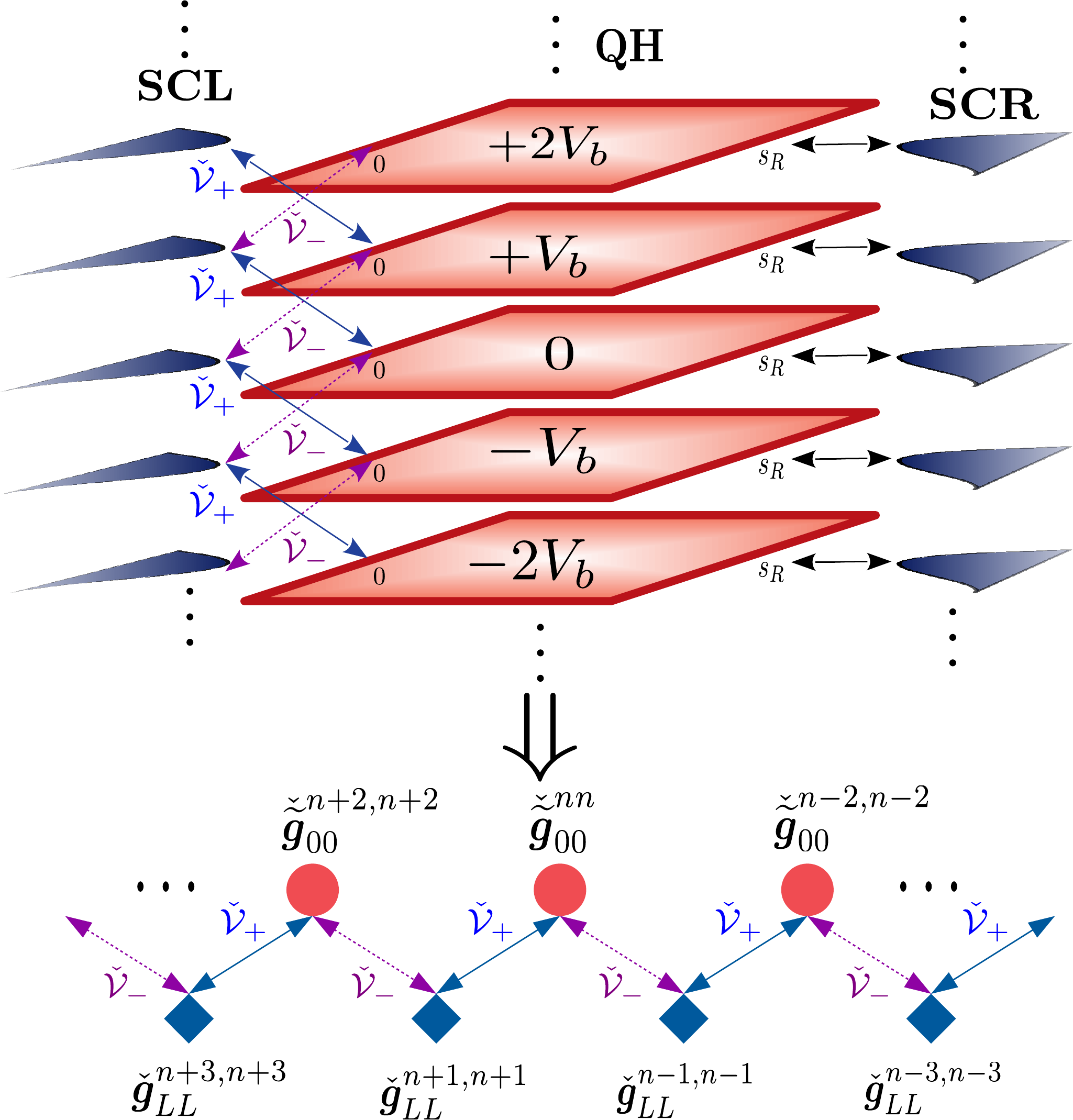}
\caption{
Top: a full scheme of the Floquet replicas of the device. They are shifted in multiples of the driving frequency $\Omega=eV_b/h$. The arrows indicate the hopping matrices between the leads and the Hall sample: $\check{\mathcal{V}}_{+}$ transfers electronic states, and $\check{\mathcal{V}}_{-}$ transfers holelike states. The right lead possesses a diagonal coupling in Floquet space due to the gauge transformation that eliminated its time dependence. Bottom: the effective one-dimensional chains (one for even $n$ and another for odd $n$) that are finally numerically solved.}
\label{fig_app}
\end{figure}
One should bear in mind that the Floquet representation of any equilibrium Green's function $\bm{g}(\omega)$ has a block-diagonal form such that $\check{\bm{g}}^{mn}(\omega)=\delta_{m,n}\,\check{g}(\omega+n\Omega)$, where $\check{g}(\omega)$ is the usual Fourier transform of a time-translationally invariant propagator written in the Nambu basis. In particular, the uncoupled lesser Green's functions will bear Fermi-Dirac distributions shifted in multiples of the frequency, such that
\begin{equation}
\check{\bm{g}}^{< nn}(\omega)=f(\omega+n\Omega)\left[\check{g}^{a}(\omega+n\Omega)-\check{g}^{r}(\omega+n\Omega)\right]\,,
\end{equation}
with $\check{g}^{a/r}(\omega)$ being the corresponding advanced and retarded Green's functions. We will focus on the zero temperature regime, so that $f(\omega)=\Theta(-\omega)$, with $\Theta(\omega)$ being the Heaviside function.

In this case, the equilibrium Green's functions of the superconducting leads are written in Nambu space as
\begin{equation}
\check{g}^{r/a}_{\nu\nu}(\omega)=\frac{\pi\rho(\epsilon_F)}{\sqrt{\Delta^2-(\omega\pm i\eta)^2}}\begin{pmatrix}
-(\omega\pm i\eta) & \Delta\\
\Delta & -(\omega\pm i\eta)
\end{pmatrix},
\end{equation}
where $\rho(\epsilon_F)$ is the leads normal density of states at the Fermi energy and finite bandwidth effects have been neglected. On the other hand, the equilibrium (uncoupled) Green's functions of the chiral state, which will be ultimately needed to obtain the current, are found to be

\begin{widetext}
\begin{equation}
\check{g}_{00}^{r/a}(\omega)=\check{g}_{\alpha\alpha}^{r/a}(\omega)=\frac{\pi}{\Lambda \varepsilon_0}
\begin{pmatrix}
\cot\left(\pi\frac{\omega + e V_b/2 \pm i\eta}{\varepsilon_0}+\pi\frac{\Phi}{\Phi_0}\right) & 0\\
0 & \cot\left(\pi\frac{\omega - e V_b/2 \pm i\eta}{\varepsilon_0}-\pi\frac{\Phi}{\Phi_0}\right)\,,
\end{pmatrix}
\end{equation}
and
\begin{equation}
\check{g}_{0\alpha}^{r/a}(\omega)=\frac{\pi}{\Lambda \varepsilon_0}
\begin{pmatrix}
e^{i(\pi-\alpha)\left(\frac{\omega+eV_b/2\pm i\eta}{\varepsilon_0}+\frac{\Phi}{\Phi_0}\right)}\csc\left(\pi\frac{\omega + e V_b/2 \pm i\eta}{\varepsilon_0}+\pi\frac{\Phi}{\Phi_0}\right) & 0\\
0 & e^{i(\pi-\alpha)\left(\frac{\omega-eV_b/2\pm i\eta}{\varepsilon_0}-\frac{\Phi}{\Phi_0}\right)}\csc\left(\pi\frac{\omega - e V_b/2 \pm i\eta}{\varepsilon_0}-\pi\frac{\Phi}{\Phi_0}\right)
\end{pmatrix}\, .
\end{equation}
\end{widetext}

Here $\check{g}_{00}(\omega)$ and $\check{g}_{\alpha\alpha}(\omega)$ are the local propagators at the sites which are coupled to the left and right lead, respectively. The nonlocal propagator $\check{g}_{0\alpha}(\omega)$ goes from site $s=0$ to the one at site $s_R = \Lambda\alpha/2\pi$, while $\check{g}_{\alpha 0}(\omega)$ may be obtained by changing $\alpha\rightarrow 2\pi-\alpha$ in $\check{g}_{0\alpha}(\omega)$.

To better illustrate the method, we show in Fig.~\ref{fig_app} a pictorial scheme of the Floquet replicas of the device, where the nondiagonal couplings $\check{\mathcal{V}}_{\pm}$ defined in Eq.~\eqref{floquet_hoppings} are only present between the left superconducting lead and the $s=0$ site of the Hall edge. The right lead, located at $s_R = \Lambda \alpha/2\pi$, has a diagonal coupling in the Floquet representation due to our choice of gauge [see Eq.~\eqref{gauge}], which eliminated the time dependence in this link. This makes feasible a procedure where both the right lead and the entire perimeter of the Hall bar are included in an effective Green's function of the zeroth site, which is given by
\begin{equation}
\check{\widetilde{g}}^{r/a}_{00}(\omega) = \check{g}_{00}^{r/a}(\omega) + \check{g}_{0\alpha}^{r/a}(\omega)\check{\Sigma}^{r/a}_{\alpha\alpha}(\omega)\check{g}_{\alpha 0}^{r/a}(\omega),
\end{equation}
where we have defined the self-energy
\begin{eqnarray}
\check{\Sigma}_{\alpha\alpha}^{r/a}(\omega)&=&\check{\mathcal{V}}_{\alpha R}[\check{g}^{r/a-1}_{RR}(\omega)-\check{\Sigma}_{RR}^{r/a}(\omega)]^{-1}\check{\mathcal{V}}_{R\alpha}\\
\notag
\check{\Sigma}_{RR}^{r/a}(\omega)&=&\check{\mathcal{V}}_{R\alpha}\check{g}^{r/a}_{\alpha\alpha}(\omega)\check{\mathcal{V}}_{\alpha R},
\end{eqnarray}
and the hopping $\check{\mathcal{V}}_{\alpha R} = \check{\mathcal{V}}_{R\alpha} = -\gamma\tau_z$. The corresponding lesser Green's function is obtained as
\begin{equation}
\check{\widetilde{g}}^{<}_{00}(\omega) = \check{\widetilde{g}}^{r}_{0\alpha}(\omega)\check{\Sigma}^{<}_{\alpha\alpha}(\omega)\check{\widetilde{g}}^{a}_{\alpha 0}(\omega),
\end{equation}
with $\check{\Sigma}^{<}_{\alpha\alpha} = \check{\mathcal{V}}_{\alpha R}\check{g}^{<}_{RR}(\omega)\check{\mathcal{V}}_{R\alpha}$.

The Floquet replicas of these effective sites $\check{\widetilde{\bm{g}}}_{00}^{nn}=\check{\widetilde{g}}_{00}(\omega+n\Omega)$ are depicted with circles in the lower panel of Fig.~\ref{fig_app}. We also indicate their coupling to the replicas of the left superconducting lead, shown as diamonds. The problem has then been reduced to solving two such one-dimensional chains, one for even and another for odd $n$.
The Floquet-Dyson equations of motion, written in an infinite-dimensional matrix representation, are finally given by
\begin{equation}
\bm{G}^{<}_{0L}(\omega)=\bm{G}^{r}_{00}(\omega)\bm{\mathcal{V}}_{0L}\bm{g}^{<}_{LL}+\bm{G}^{<}_{00}(\omega)\bm{\mathcal{V}}_{0L}\bm{g}^{a}_{LL},
\end{equation} 
with
\begin{eqnarray}
\bm{G}^{r}_{00} &=& \widetilde{\bm{g}}^{r}_{00}+\widetilde{\bm{g}}^{r}_{00}\bm{\Sigma}^{r}_{00}\bm{G}^{r}_{00}\\
\notag
\bm{G}^{<}_{00}&=&\bm{G}^{r}_{00}\bm{\Sigma}^{<}_{00}\bm{G}^{a}_{00}+(\mathbb{1} +\bm{G}^{r}_{00}\bm{\Sigma}^{r}_{00})\widetilde{\bm{g}}^{<}_{00}(\mathbb{1} +\bm{\Sigma}^{a}_{00}\bm{G}^{a}_{00})\, ,
\end{eqnarray}
and where $\bm{\Sigma}^{r/a/<}_{00}(\omega) = \bm{\mathcal{V}}_{0L}\bm{g}_{LL}^{r/a/<}(\omega)\bm{\mathcal{V}}_{L0}$. The block elements of $\bm{\mathcal{V}}_{L0}$ have been defined in Eq.~\eqref{floquet_hoppings} and $\bm{\mathcal{V}}_{0L}=\bm{\mathcal{V}}_{L0}^{\dagger}$. This set of equations can be solved with standard recursion techniques~\citep{Cuevas1996} by increasing the number of Floquet replicas until numerical convergence of the results is achieved.


\begin{thebibliography}{41}%
\makeatletter
\providecommand \@ifxundefined [1]{%
 \@ifx{#1\undefined}
}%
\providecommand \@ifnum [1]{%
 \ifnum #1\expandafter \@firstoftwo
 \else \expandafter \@secondoftwo
 \fi
}%
\providecommand \@ifx [1]{%
 \ifx #1\expandafter \@firstoftwo
 \else \expandafter \@secondoftwo
 \fi
}%
\providecommand \natexlab [1]{#1}%
\providecommand \enquote  [1]{``#1''}%
\providecommand \bibnamefont  [1]{#1}%
\providecommand \bibfnamefont [1]{#1}%
\providecommand \citenamefont [1]{#1}%
\providecommand \href@noop [0]{\@secondoftwo}%
\providecommand \href [0]{\begingroup \@sanitize@url \@href}%
\providecommand \@href[1]{\@@startlink{#1}\@@href}%
\providecommand \@@href[1]{\endgroup#1\@@endlink}%
\providecommand \@sanitize@url [0]{\catcode `\\12\catcode `\$12\catcode
  `\&12\catcode `\#12\catcode `\^12\catcode `\_12\catcode `\%12\relax}%
\providecommand \@@startlink[1]{}%
\providecommand \@@endlink[0]{}%
\providecommand \url  [0]{\begingroup\@sanitize@url \@url }%
\providecommand \@url [1]{\endgroup\@href {#1}{\urlprefix }}%
\providecommand \urlprefix  [0]{URL }%
\providecommand \Eprint [0]{\href }%
\providecommand \doibase [0]{https://doi.org/}%
\providecommand \selectlanguage [0]{\@gobble}%
\providecommand \bibinfo  [0]{\@secondoftwo}%
\providecommand \bibfield  [0]{\@secondoftwo}%
\providecommand \translation [1]{[#1]}%
\providecommand \BibitemOpen [0]{}%
\providecommand \bibitemStop [0]{}%
\providecommand \bibitemNoStop [0]{.\EOS\space}%
\providecommand \EOS [0]{\spacefactor3000\relax}%
\providecommand \BibitemShut  [1]{\csname bibitem#1\endcsname}%
\let\auto@bib@innerbib\@empty
\bibitem [{\citenamefont {Hoppe}\ \emph {et~al.}(2000)\citenamefont {Hoppe},
  \citenamefont {Z\"ulicke},\ and\ \citenamefont {Sch\"on}}]{Hoppe2000}%
  \BibitemOpen
  \bibfield  {author} {\bibinfo {author} {\bibfnamefont {H.}~\bibnamefont
  {Hoppe}}, \bibinfo {author} {\bibfnamefont {U.}~\bibnamefont {Z\"ulicke}},\
  and\ \bibinfo {author} {\bibfnamefont {G.}~\bibnamefont {Sch\"on}},\
  }\bibfield  {title} {\bibinfo {title} {Andreev {R}eflection in {S}trong
  {M}agnetic {F}ields},\ }\href {https://doi.org/10.1103/PhysRevLett.84.1804}
  {\bibfield  {journal} {\bibinfo  {journal} {Phys. Rev. Lett.}\ }\textbf
  {\bibinfo {volume} {84}},\ \bibinfo {pages} {1804} (\bibinfo {year}
  {2000})}\BibitemShut {NoStop}%
\bibitem [{\citenamefont {Giazotto}\ \emph {et~al.}(2005)\citenamefont
  {Giazotto}, \citenamefont {Governale}, \citenamefont {Z\"ulicke},\ and\
  \citenamefont {Beltram}}]{Giazotto2005}%
  \BibitemOpen
  \bibfield  {author} {\bibinfo {author} {\bibfnamefont {F.}~\bibnamefont
  {Giazotto}}, \bibinfo {author} {\bibfnamefont {M.}~\bibnamefont {Governale}},
  \bibinfo {author} {\bibfnamefont {U.}~\bibnamefont {Z\"ulicke}},\ and\
  \bibinfo {author} {\bibfnamefont {F.}~\bibnamefont {Beltram}},\ }\bibfield
  {title} {\bibinfo {title} {Andreev reflection and cyclotron motion at
  superconductor---normal-metal interfaces},\ }\href
  {https://doi.org/10.1103/PhysRevB.72.054518} {\bibfield  {journal} {\bibinfo
  {journal} {Phys. Rev. B}\ }\textbf {\bibinfo {volume} {72}},\ \bibinfo
  {pages} {054518} (\bibinfo {year} {2005})}\BibitemShut {NoStop}%
\bibitem [{\citenamefont {Hou}\ \emph {et~al.}(2016)\citenamefont {Hou},
  \citenamefont {Xing}, \citenamefont {Guo},\ and\ \citenamefont
  {Sun}}]{Hou2016}%
  \BibitemOpen
  \bibfield  {author} {\bibinfo {author} {\bibfnamefont {Z.}~\bibnamefont
  {Hou}}, \bibinfo {author} {\bibfnamefont {Y.}~\bibnamefont {Xing}}, \bibinfo
  {author} {\bibfnamefont {A.-M.}\ \bibnamefont {Guo}},\ and\ \bibinfo {author}
  {\bibfnamefont {Q.-F.}\ \bibnamefont {Sun}},\ }\bibfield  {title} {\bibinfo
  {title} {Crossed {Andreev} effects in two-dimensional quantum {Hall}
  systems},\ }\href {https://doi.org/10.1103/PhysRevB.94.064516} {\bibfield
  {journal} {\bibinfo  {journal} {Physical Review B}\ }\textbf {\bibinfo
  {volume} {94}},\ \bibinfo {pages} {064516} (\bibinfo {year}
  {2016})}\BibitemShut {NoStop}%
\bibitem [{\citenamefont {Lee}\ \emph {et~al.}(2017)\citenamefont {Lee},
  \citenamefont {Huang}, \citenamefont {Efetov}, \citenamefont {Wei},
  \citenamefont {Hart}, \citenamefont {Taniguchi}, \citenamefont {Watanabe},
  \citenamefont {Yacoby},\ and\ \citenamefont {Kim}}]{Lee2017}%
  \BibitemOpen
  \bibfield  {author} {\bibinfo {author} {\bibfnamefont {G.-H.}\ \bibnamefont
  {Lee}}, \bibinfo {author} {\bibfnamefont {K.-F.}\ \bibnamefont {Huang}},
  \bibinfo {author} {\bibfnamefont {D.~K.}\ \bibnamefont {Efetov}}, \bibinfo
  {author} {\bibfnamefont {D.~S.}\ \bibnamefont {Wei}}, \bibinfo {author}
  {\bibfnamefont {S.}~\bibnamefont {Hart}}, \bibinfo {author} {\bibfnamefont
  {T.}~\bibnamefont {Taniguchi}}, \bibinfo {author} {\bibfnamefont
  {K.}~\bibnamefont {Watanabe}}, \bibinfo {author} {\bibfnamefont
  {A.}~\bibnamefont {Yacoby}},\ and\ \bibinfo {author} {\bibfnamefont
  {P.}~\bibnamefont {Kim}},\ }\bibfield  {title} {\bibinfo {title} {Inducing
  superconducting correlation in quantum {H}all edge states},\ }\href
  {https://doi.org/10.1038/nphys4084} {\bibfield  {journal} {\bibinfo
  {journal} {Nature Physics}\ }\textbf {\bibinfo {volume} {13}},\ \bibinfo
  {pages} {693} (\bibinfo {year} {2017})}\BibitemShut {NoStop}%
\bibitem [{\citenamefont {Lindner}\ \emph {et~al.}(2012)\citenamefont
  {Lindner}, \citenamefont {Berg}, \citenamefont {Refael},\ and\ \citenamefont
  {Stern}}]{Lindner2012}%
  \BibitemOpen
  \bibfield  {author} {\bibinfo {author} {\bibfnamefont {N.~H.}\ \bibnamefont
  {Lindner}}, \bibinfo {author} {\bibfnamefont {E.}~\bibnamefont {Berg}},
  \bibinfo {author} {\bibfnamefont {G.}~\bibnamefont {Refael}},\ and\ \bibinfo
  {author} {\bibfnamefont {A.}~\bibnamefont {Stern}},\ }\bibfield  {title}
  {\bibinfo {title} {Fractionalizing {Majorana} {F}ermions: {N}on-{Abelian}
  {S}tatistics on the {E}dges of {Abelian} {Q}uantum {Hall} {S}tates},\ }\href
  {https://doi.org/10.1103/PhysRevX.2.041002} {\bibfield  {journal} {\bibinfo
  {journal} {Phys. Rev. X}\ }\textbf {\bibinfo {volume} {2}},\ \bibinfo {pages}
  {041002} (\bibinfo {year} {2012})}\BibitemShut {NoStop}%
\bibitem [{\citenamefont {Clarke}\ \emph {et~al.}(2013)\citenamefont {Clarke},
  \citenamefont {Alicea},\ and\ \citenamefont {Shtengel}}]{Clarke2013}%
  \BibitemOpen
  \bibfield  {author} {\bibinfo {author} {\bibfnamefont {D.~J.}\ \bibnamefont
  {Clarke}}, \bibinfo {author} {\bibfnamefont {J.}~\bibnamefont {Alicea}},\
  and\ \bibinfo {author} {\bibfnamefont {K.}~\bibnamefont {Shtengel}},\
  }\bibfield  {title} {\bibinfo {title} {Exotic non-{Abelian} anyons from
  conventional fractional quantum {Hall} states},\ }\href
  {https://doi.org/10.1038/ncomms2340} {\bibfield  {journal} {\bibinfo
  {journal} {Nature Communications}\ }\textbf {\bibinfo {volume} {4}},\
  \bibinfo {pages} {1348} (\bibinfo {year} {2013})}\BibitemShut {NoStop}%
\bibitem [{\citenamefont {Qi}\ \emph {et~al.}(2010)\citenamefont {Qi},
  \citenamefont {Hughes},\ and\ \citenamefont {Zhang}}]{Qi2010}%
  \BibitemOpen
  \bibfield  {author} {\bibinfo {author} {\bibfnamefont {X.-L.}\ \bibnamefont
  {Qi}}, \bibinfo {author} {\bibfnamefont {T.~L.}\ \bibnamefont {Hughes}},\
  and\ \bibinfo {author} {\bibfnamefont {S.-C.}\ \bibnamefont {Zhang}},\
  }\bibfield  {title} {\bibinfo {title} {Chiral topological superconductor from
  the quantum {Hall} state},\ }\href
  {https://doi.org/10.1103/PhysRevB.82.184516} {\bibfield  {journal} {\bibinfo
  {journal} {Phys. Rev. B}\ }\textbf {\bibinfo {volume} {82}},\ \bibinfo
  {pages} {184516} (\bibinfo {year} {2010})}\BibitemShut {NoStop}%
\bibitem [{\citenamefont {Chaudhary}\ and\ \citenamefont
  {MacDonald}(2020)}]{Chaudhary2020}%
  \BibitemOpen
  \bibfield  {author} {\bibinfo {author} {\bibfnamefont {G.}~\bibnamefont
  {Chaudhary}}\ and\ \bibinfo {author} {\bibfnamefont {A.~H.}\ \bibnamefont
  {MacDonald}},\ }\bibfield  {title} {\bibinfo {title} {Vortex-lattice
  structure and topological superconductivity in the quantum {Hall} regime},\
  }\href {https://doi.org/10.1103/PhysRevB.101.024516} {\bibfield  {journal}
  {\bibinfo  {journal} {Physical Review B}\ }\textbf {\bibinfo {volume}
  {101}},\ \bibinfo {pages} {024516} (\bibinfo {year} {2020})}\BibitemShut
  {NoStop}%
\bibitem [{\citenamefont {Wan}\ \emph {et~al.}(2015)\citenamefont {Wan},
  \citenamefont {Kazakov}, \citenamefont {Manfra}, \citenamefont {Pfeiffer},
  \citenamefont {West},\ and\ \citenamefont {Rokhinson}}]{Wan2015}%
  \BibitemOpen
  \bibfield  {author} {\bibinfo {author} {\bibfnamefont {Z.}~\bibnamefont
  {Wan}}, \bibinfo {author} {\bibfnamefont {A.}~\bibnamefont {Kazakov}},
  \bibinfo {author} {\bibfnamefont {M.~J.}\ \bibnamefont {Manfra}}, \bibinfo
  {author} {\bibfnamefont {L.~N.}\ \bibnamefont {Pfeiffer}}, \bibinfo {author}
  {\bibfnamefont {K.~W.}\ \bibnamefont {West}},\ and\ \bibinfo {author}
  {\bibfnamefont {L.~P.}\ \bibnamefont {Rokhinson}},\ }\bibfield  {title}
  {\bibinfo {title} {Induced superconductivity in high-mobility two-dimensional
  electron gas in gallium arsenide heterostructures},\ }\href
  {https://doi.org/10.1038/ncomms8426} {\bibfield  {journal} {\bibinfo
  {journal} {Nature Communications}\ }\textbf {\bibinfo {volume} {6}},\
  \bibinfo {pages} {7426} (\bibinfo {year} {2015})}\BibitemShut {NoStop}%
\bibitem [{\citenamefont {Amet}\ \emph {et~al.}(2016)\citenamefont {Amet},
  \citenamefont {Ke}, \citenamefont {Borzenets}, \citenamefont {Wang},
  \citenamefont {Watanabe}, \citenamefont {Taniguchi}, \citenamefont {Deacon},
  \citenamefont {Yamamoto}, \citenamefont {Bomze}, \citenamefont {Tarucha},\
  and\ \citenamefont {Finkelstein}}]{Amet2016}%
  \BibitemOpen
  \bibfield  {author} {\bibinfo {author} {\bibfnamefont {F.}~\bibnamefont
  {Amet}}, \bibinfo {author} {\bibfnamefont {C.~T.}\ \bibnamefont {Ke}},
  \bibinfo {author} {\bibfnamefont {I.~V.}\ \bibnamefont {Borzenets}}, \bibinfo
  {author} {\bibfnamefont {J.}~\bibnamefont {Wang}}, \bibinfo {author}
  {\bibfnamefont {K.}~\bibnamefont {Watanabe}}, \bibinfo {author}
  {\bibfnamefont {T.}~\bibnamefont {Taniguchi}}, \bibinfo {author}
  {\bibfnamefont {R.~S.}\ \bibnamefont {Deacon}}, \bibinfo {author}
  {\bibfnamefont {M.}~\bibnamefont {Yamamoto}}, \bibinfo {author}
  {\bibfnamefont {Y.}~\bibnamefont {Bomze}}, \bibinfo {author} {\bibfnamefont
  {S.}~\bibnamefont {Tarucha}},\ and\ \bibinfo {author} {\bibfnamefont
  {G.}~\bibnamefont {Finkelstein}},\ }\bibfield  {title} {\bibinfo {title}
  {Supercurrent in the quantum {H}all regime},\ }\href
  {https://doi.org/10.1126/science.aad6203} {\bibfield  {journal} {\bibinfo
  {journal} {Science}\ }\textbf {\bibinfo {volume} {352}},\ \bibinfo {pages}
  {966} (\bibinfo {year} {2016})}\BibitemShut {NoStop}%
\bibitem [{\citenamefont {Guiducci}\ \emph {et~al.}(2018)\citenamefont
  {Guiducci}, \citenamefont {Carrega}, \citenamefont {Biasiol}, \citenamefont
  {Sorba}, \citenamefont {Beltram},\ and\ \citenamefont {Heun}}]{Guiducci2018}%
  \BibitemOpen
  \bibfield  {author} {\bibinfo {author} {\bibfnamefont {S.}~\bibnamefont
  {Guiducci}}, \bibinfo {author} {\bibfnamefont {M.}~\bibnamefont {Carrega}},
  \bibinfo {author} {\bibfnamefont {G.}~\bibnamefont {Biasiol}}, \bibinfo
  {author} {\bibfnamefont {L.}~\bibnamefont {Sorba}}, \bibinfo {author}
  {\bibfnamefont {F.}~\bibnamefont {Beltram}},\ and\ \bibinfo {author}
  {\bibfnamefont {S.}~\bibnamefont {Heun}},\ }\bibfield  {title} {\bibinfo
  {title} {Toward quantum {H}all effect in a {J}osephson junction},\ }\href
  {https://doi.org/10.1002/pssr.201800222} {\bibfield  {journal} {\bibinfo
  {journal} {Physica Status Solidi ({RRL})}\ }\textbf {\bibinfo {volume}
  {13}},\ \bibinfo {pages} {1800222} (\bibinfo {year} {2018})}\BibitemShut
  {NoStop}%
\bibitem [{\citenamefont {Seredinski}\ \emph {et~al.}(2019)\citenamefont
  {Seredinski}, \citenamefont {Draelos}, \citenamefont {Arnault}, \citenamefont
  {Wei}, \citenamefont {Li}, \citenamefont {Fleming}, \citenamefont {Watanabe},
  \citenamefont {Taniguchi}, \citenamefont {Amet},\ and\ \citenamefont
  {Finkelstein}}]{Seredinski2019}%
  \BibitemOpen
  \bibfield  {author} {\bibinfo {author} {\bibfnamefont {A.}~\bibnamefont
  {Seredinski}}, \bibinfo {author} {\bibfnamefont {A.~W.}\ \bibnamefont
  {Draelos}}, \bibinfo {author} {\bibfnamefont {E.~G.}\ \bibnamefont
  {Arnault}}, \bibinfo {author} {\bibfnamefont {M.-T.}\ \bibnamefont {Wei}},
  \bibinfo {author} {\bibfnamefont {H.}~\bibnamefont {Li}}, \bibinfo {author}
  {\bibfnamefont {T.}~\bibnamefont {Fleming}}, \bibinfo {author} {\bibfnamefont
  {K.}~\bibnamefont {Watanabe}}, \bibinfo {author} {\bibfnamefont
  {T.}~\bibnamefont {Taniguchi}}, \bibinfo {author} {\bibfnamefont
  {F.}~\bibnamefont {Amet}},\ and\ \bibinfo {author} {\bibfnamefont
  {G.}~\bibnamefont {Finkelstein}},\ }\bibfield  {title} {\bibinfo {title}
  {Quantum {H}all{\textendash}based superconducting interference device},\
  }\href {https://doi.org/10.1126/sciadv.aaw8693} {\bibfield  {journal}
  {\bibinfo  {journal} {Science Advances}\ }\textbf {\bibinfo {volume} {5}},\
  \bibinfo {pages} {eaaw8693} (\bibinfo {year} {2019})}\BibitemShut {NoStop}%
\bibitem [{\citenamefont {Zhi}\ \emph {et~al.}(2019)\citenamefont {Zhi},
  \citenamefont {Kang}, \citenamefont {Su}, \citenamefont {Fan}, \citenamefont
  {Li}, \citenamefont {Pan}, \citenamefont {Zhao}, \citenamefont {Zhao},\ and\
  \citenamefont {Xu}}]{Zhi2019}%
  \BibitemOpen
  \bibfield  {author} {\bibinfo {author} {\bibfnamefont {J.}~\bibnamefont
  {Zhi}}, \bibinfo {author} {\bibfnamefont {N.}~\bibnamefont {Kang}}, \bibinfo
  {author} {\bibfnamefont {F.}~\bibnamefont {Su}}, \bibinfo {author}
  {\bibfnamefont {D.}~\bibnamefont {Fan}}, \bibinfo {author} {\bibfnamefont
  {S.}~\bibnamefont {Li}}, \bibinfo {author} {\bibfnamefont {D.}~\bibnamefont
  {Pan}}, \bibinfo {author} {\bibfnamefont {S.~P.}\ \bibnamefont {Zhao}},
  \bibinfo {author} {\bibfnamefont {J.}~\bibnamefont {Zhao}},\ and\ \bibinfo
  {author} {\bibfnamefont {H.~Q.}\ \bibnamefont {Xu}},\ }\bibfield  {title}
  {\bibinfo {title} {Coexistence of induced superconductivity and quantum
  {Hall} states in {InSb} nanosheets},\ }\href
  {https://doi.org/10.1103/PhysRevB.99.245302} {\bibfield  {journal} {\bibinfo
  {journal} {Phys. Rev. B}\ }\textbf {\bibinfo {volume} {99}},\ \bibinfo
  {pages} {245302} (\bibinfo {year} {2019})}\BibitemShut {NoStop}%
\bibitem [{\citenamefont {Zhao}\ \emph {et~al.}(2020)\citenamefont {Zhao},
  \citenamefont {Arnault}, \citenamefont {Bondarev}, \citenamefont
  {Seredinski}, \citenamefont {Larson}, \citenamefont {Draelos}, \citenamefont
  {Li}, \citenamefont {Watanabe}, \citenamefont {Taniguchi}, \citenamefont
  {Amet}, \citenamefont {Baranger},\ and\ \citenamefont
  {Finkelstein}}]{Zhao2020}%
  \BibitemOpen
  \bibfield  {author} {\bibinfo {author} {\bibfnamefont {L.}~\bibnamefont
  {Zhao}}, \bibinfo {author} {\bibfnamefont {E.~G.}\ \bibnamefont {Arnault}},
  \bibinfo {author} {\bibfnamefont {A.}~\bibnamefont {Bondarev}}, \bibinfo
  {author} {\bibfnamefont {A.}~\bibnamefont {Seredinski}}, \bibinfo {author}
  {\bibfnamefont {T.~F.~Q.}\ \bibnamefont {Larson}}, \bibinfo {author}
  {\bibfnamefont {A.~W.}\ \bibnamefont {Draelos}}, \bibinfo {author}
  {\bibfnamefont {H.}~\bibnamefont {Li}}, \bibinfo {author} {\bibfnamefont
  {K.}~\bibnamefont {Watanabe}}, \bibinfo {author} {\bibfnamefont
  {T.}~\bibnamefont {Taniguchi}}, \bibinfo {author} {\bibfnamefont
  {F.}~\bibnamefont {Amet}}, \bibinfo {author} {\bibfnamefont {H.~U.}\
  \bibnamefont {Baranger}},\ and\ \bibinfo {author} {\bibfnamefont
  {G.}~\bibnamefont {Finkelstein}},\ }\bibfield  {title} {\bibinfo {title}
  {Interference of chiral {Andreev} edge states},\ }\href
  {https://doi.org/10.1038/s41567-020-0898-5} {\bibfield  {journal} {\bibinfo
  {journal} {Nature Physics}\ }\textbf {\bibinfo {volume} {16}},\ \bibinfo
  {pages} {862} (\bibinfo {year} {2020})}\BibitemShut {NoStop}%
\bibitem [{\citenamefont {Bhandari}\ \emph {et~al.}(2020)\citenamefont
  {Bhandari}, \citenamefont {Lee}, \citenamefont {Watanabe}, \citenamefont
  {Taniguchi}, \citenamefont {Kim},\ and\ \citenamefont
  {Westervelt}}]{Bhandari2020}%
  \BibitemOpen
  \bibfield  {author} {\bibinfo {author} {\bibfnamefont {S.}~\bibnamefont
  {Bhandari}}, \bibinfo {author} {\bibfnamefont {G.-H.}\ \bibnamefont {Lee}},
  \bibinfo {author} {\bibfnamefont {K.}~\bibnamefont {Watanabe}}, \bibinfo
  {author} {\bibfnamefont {T.}~\bibnamefont {Taniguchi}}, \bibinfo {author}
  {\bibfnamefont {P.}~\bibnamefont {Kim}},\ and\ \bibinfo {author}
  {\bibfnamefont {R.~M.}\ \bibnamefont {Westervelt}},\ }\bibfield  {title}
  {\bibinfo {title} {Imaging {Andreev} reflection in graphene},\ }\href
  {https://doi.org/10.1021/acs.nanolett.0c00903} {\bibfield  {journal}
  {\bibinfo  {journal} {Nano Letters}\ }\textbf {\bibinfo {volume} {20}},\
  \bibinfo {pages} {4890} (\bibinfo {year} {2020})}\BibitemShut {NoStop}%
\bibitem [{\citenamefont {Ma}\ and\ \citenamefont {Zyuzin}(1993)}]{Ma1993}%
  \BibitemOpen
  \bibfield  {author} {\bibinfo {author} {\bibfnamefont {M.}~\bibnamefont
  {Ma}}\ and\ \bibinfo {author} {\bibfnamefont {A.~Y.}\ \bibnamefont
  {Zyuzin}},\ }\bibfield  {title} {\bibinfo {title} {Josephson effect in the
  quantum {H}all regime},\ }\href {https://doi.org/10.1209/0295-5075/21/9/011}
  {\bibfield  {journal} {\bibinfo  {journal} {Europhysics Letters ({EPL})}\
  }\textbf {\bibinfo {volume} {21}},\ \bibinfo {pages} {941} (\bibinfo {year}
  {1993})}\BibitemShut {NoStop}%
\bibitem [{\citenamefont {van Ostaay}\ \emph {et~al.}(2011)\citenamefont {van
  Ostaay}, \citenamefont {Akhmerov},\ and\ \citenamefont
  {Beenakker}}]{vanOstaay2011}%
  \BibitemOpen
  \bibfield  {author} {\bibinfo {author} {\bibfnamefont {J.~A.~M.}\
  \bibnamefont {van Ostaay}}, \bibinfo {author} {\bibfnamefont {A.~R.}\
  \bibnamefont {Akhmerov}},\ and\ \bibinfo {author} {\bibfnamefont {C.~W.~J.}\
  \bibnamefont {Beenakker}},\ }\bibfield  {title} {\bibinfo {title}
  {Spin-triplet supercurrent carried by quantum {H}all edge states through a
  {J}osephson junction},\ }\href {https://doi.org/10.1103/PhysRevB.83.195441}
  {\bibfield  {journal} {\bibinfo  {journal} {Phys. Rev. B}\ }\textbf {\bibinfo
  {volume} {83}},\ \bibinfo {pages} {195441} (\bibinfo {year}
  {2011})}\BibitemShut {NoStop}%
\bibitem [{\citenamefont {Stone}\ and\ \citenamefont {Lin}(2011)}]{Stone2011}%
  \BibitemOpen
  \bibfield  {author} {\bibinfo {author} {\bibfnamefont {M.}~\bibnamefont
  {Stone}}\ and\ \bibinfo {author} {\bibfnamefont {Y.}~\bibnamefont {Lin}},\
  }\bibfield  {title} {\bibinfo {title} {Josephson currents in quantum {H}all
  devices},\ }\href {https://doi.org/10.1103/PhysRevB.83.224501} {\bibfield
  {journal} {\bibinfo  {journal} {Phys. Rev. B}\ }\textbf {\bibinfo {volume}
  {83}},\ \bibinfo {pages} {224501} (\bibinfo {year} {2011})}\BibitemShut
  {NoStop}%
\bibitem [{\citenamefont {Alavirad}\ \emph {et~al.}(2018)\citenamefont
  {Alavirad}, \citenamefont {Lee}, \citenamefont {Lin},\ and\ \citenamefont
  {Sau}}]{Alavirad2018}%
  \BibitemOpen
  \bibfield  {author} {\bibinfo {author} {\bibfnamefont {Y.}~\bibnamefont
  {Alavirad}}, \bibinfo {author} {\bibfnamefont {J.}~\bibnamefont {Lee}},
  \bibinfo {author} {\bibfnamefont {Z.-X.}\ \bibnamefont {Lin}},\ and\ \bibinfo
  {author} {\bibfnamefont {J.~D.}\ \bibnamefont {Sau}},\ }\bibfield  {title}
  {\bibinfo {title} {Chiral supercurrent through a quantum {H}all weak link},\
  }\href {https://doi.org/10.1103/PhysRevB.98.214504} {\bibfield  {journal}
  {\bibinfo  {journal} {Phys. Rev. B}\ }\textbf {\bibinfo {volume} {98}},\
  \bibinfo {pages} {214504} (\bibinfo {year} {2018})}\BibitemShut {NoStop}%
\bibitem [{\citenamefont {Peralta~Gavensky}\ \emph {et~al.}(2020)\citenamefont
  {Peralta~Gavensky}, \citenamefont {Usaj},\ and\ \citenamefont
  {Balseiro}}]{PeraltaGavensky2020}%
  \BibitemOpen
  \bibfield  {author} {\bibinfo {author} {\bibfnamefont {L.}~\bibnamefont
  {Peralta~Gavensky}}, \bibinfo {author} {\bibfnamefont {G.}~\bibnamefont
  {Usaj}},\ and\ \bibinfo {author} {\bibfnamefont {C.~A.}\ \bibnamefont
  {Balseiro}},\ }\bibfield  {title} {\bibinfo {title} {Majorana fermions on the
  quantum {Hall} edge},\ }\href
  {https://doi.org/10.1103/PhysRevResearch.2.033218} {\bibfield  {journal}
  {\bibinfo  {journal} {Phys. Rev. Research}\ }\textbf {\bibinfo {volume}
  {2}},\ \bibinfo {pages} {033218} (\bibinfo {year} {2020})}\BibitemShut
  {NoStop}%
\bibitem [{\citenamefont {Octavio}\ \emph {et~al.}(1983)\citenamefont
  {Octavio}, \citenamefont {Tinkham}, \citenamefont {Blonder},\ and\
  \citenamefont {Klapwijk}}]{Klapwijk1983}%
  \BibitemOpen
  \bibfield  {author} {\bibinfo {author} {\bibfnamefont {M.}~\bibnamefont
  {Octavio}}, \bibinfo {author} {\bibfnamefont {M.}~\bibnamefont {Tinkham}},
  \bibinfo {author} {\bibfnamefont {G.~E.}\ \bibnamefont {Blonder}},\ and\
  \bibinfo {author} {\bibfnamefont {T.~M.}\ \bibnamefont {Klapwijk}},\
  }\bibfield  {title} {\bibinfo {title} {Subharmonic energy-gap structure in
  superconducting constrictions},\ }\href
  {https://doi.org/10.1103/PhysRevB.27.6739} {\bibfield  {journal} {\bibinfo
  {journal} {Phys. Rev. B}\ }\textbf {\bibinfo {volume} {27}},\ \bibinfo
  {pages} {6739} (\bibinfo {year} {1983})}\BibitemShut {NoStop}%
\bibitem [{\citenamefont {Bratus'}\ \emph {et~al.}(1995)\citenamefont
  {Bratus'}, \citenamefont {Shumeiko},\ and\ \citenamefont
  {Wendin}}]{Bratus1995}%
  \BibitemOpen
  \bibfield  {author} {\bibinfo {author} {\bibfnamefont {E.~N.}\ \bibnamefont
  {Bratus'}}, \bibinfo {author} {\bibfnamefont {V.~S.}\ \bibnamefont
  {Shumeiko}},\ and\ \bibinfo {author} {\bibfnamefont {G.}~\bibnamefont
  {Wendin}},\ }\bibfield  {title} {\bibinfo {title} {Theory of {S}ubharmonic
  {G}ap {S}tructure in {S}uperconducting {M}esoscopic {T}unnel {C}ontacts},\
  }\href {https://doi.org/10.1103/PhysRevLett.74.2110} {\bibfield  {journal}
  {\bibinfo  {journal} {Phys. Rev. Lett.}\ }\textbf {\bibinfo {volume} {74}},\
  \bibinfo {pages} {2110} (\bibinfo {year} {1995})}\BibitemShut {NoStop}%
\bibitem [{\citenamefont {Averin}\ and\ \citenamefont
  {Bardas}(1995)}]{Averin1995}%
  \BibitemOpen
  \bibfield  {author} {\bibinfo {author} {\bibfnamefont {D.}~\bibnamefont
  {Averin}}\ and\ \bibinfo {author} {\bibfnamefont {A.}~\bibnamefont
  {Bardas}},\ }\bibfield  {title} {\bibinfo {title} {ac {Josephson} {E}ffect in
  a {S}ingle {Q}uantum {C}hannel},\ }\href
  {https://doi.org/10.1103/PhysRevLett.75.1831} {\bibfield  {journal} {\bibinfo
   {journal} {Phys. Rev. Lett.}\ }\textbf {\bibinfo {volume} {75}},\ \bibinfo
  {pages} {1831} (\bibinfo {year} {1995})}\BibitemShut {NoStop}%
\bibitem [{\citenamefont {Cuevas}\ \emph {et~al.}(1996)\citenamefont {Cuevas},
  \citenamefont {Mart\'{\i}n-Rodero},\ and\ \citenamefont
  {Levy~Yeyati}}]{Cuevas1996}%
  \BibitemOpen
  \bibfield  {author} {\bibinfo {author} {\bibfnamefont {J.~C.}\ \bibnamefont
  {Cuevas}}, \bibinfo {author} {\bibfnamefont {A.}~\bibnamefont
  {Mart\'{\i}n-Rodero}},\ and\ \bibinfo {author} {\bibfnamefont
  {A.}~\bibnamefont {Levy~Yeyati}},\ }\bibfield  {title} {\bibinfo {title}
  {Hamiltonian approach to the transport properties of superconducting quantum
  point contacts},\ }\href {https://doi.org/10.1103/PhysRevB.54.7366}
  {\bibfield  {journal} {\bibinfo  {journal} {Phys. Rev. B}\ }\textbf {\bibinfo
  {volume} {54}},\ \bibinfo {pages} {7366} (\bibinfo {year}
  {1996})}\BibitemShut {NoStop}%
\bibitem [{\citenamefont {Levy~Yeyati}\ \emph {et~al.}(1997)\citenamefont
  {Levy~Yeyati}, \citenamefont {Cuevas}, \citenamefont {L\'opez-D\'avalos},\
  and\ \citenamefont {Mart\'{\i}n-Rodero}}]{Yeyati1997}%
  \BibitemOpen
  \bibfield  {author} {\bibinfo {author} {\bibfnamefont {A.}~\bibnamefont
  {Levy~Yeyati}}, \bibinfo {author} {\bibfnamefont {J.~C.}\ \bibnamefont
  {Cuevas}}, \bibinfo {author} {\bibfnamefont {A.}~\bibnamefont
  {L\'opez-D\'avalos}},\ and\ \bibinfo {author} {\bibfnamefont
  {A.}~\bibnamefont {Mart\'{\i}n-Rodero}},\ }\bibfield  {title} {\bibinfo
  {title} {Resonant tunneling through a small quantum dot coupled to
  superconducting leads},\ }\href {https://doi.org/10.1103/PhysRevB.55.R6137}
  {\bibfield  {journal} {\bibinfo  {journal} {Phys. Rev. B}\ }\textbf {\bibinfo
  {volume} {55}},\ \bibinfo {pages} {R6137} (\bibinfo {year}
  {1997})}\BibitemShut {NoStop}%
\bibitem [{\citenamefont {Johansson}\ \emph
  {et~al.}(1999{\natexlab{a}})\citenamefont {Johansson}, \citenamefont
  {Bratus}, \citenamefont {Shumeiko},\ and\ \citenamefont
  {Wendin}}]{Johansson1999}%
  \BibitemOpen
  \bibfield  {author} {\bibinfo {author} {\bibfnamefont {G.}~\bibnamefont
  {Johansson}}, \bibinfo {author} {\bibfnamefont {E.~N.}\ \bibnamefont
  {Bratus}}, \bibinfo {author} {\bibfnamefont {V.~S.}\ \bibnamefont
  {Shumeiko}},\ and\ \bibinfo {author} {\bibfnamefont {G.}~\bibnamefont
  {Wendin}},\ }\bibfield  {title} {\bibinfo {title} {Resonant multiple
  {Andreev} reflections in mesoscopic superconducting junctions},\ }\href
  {https://doi.org/10.1103/PhysRevB.60.1382} {\bibfield  {journal} {\bibinfo
  {journal} {Phys. Rev. B}\ }\textbf {\bibinfo {volume} {60}},\ \bibinfo
  {pages} {1382} (\bibinfo {year} {1999}{\natexlab{a}})}\BibitemShut {NoStop}%
\bibitem [{\citenamefont {Buitelaar}\ \emph {et~al.}(2003)\citenamefont
  {Buitelaar}, \citenamefont {Belzig}, \citenamefont {Nussbaumer},
  \citenamefont {Babi\ifmmode~\acute{c}\else \'{c}\fi{}}, \citenamefont
  {Bruder},\ and\ \citenamefont {Sch\"onenberger}}]{Buitelaar2003}%
  \BibitemOpen
  \bibfield  {author} {\bibinfo {author} {\bibfnamefont {M.~R.}\ \bibnamefont
  {Buitelaar}}, \bibinfo {author} {\bibfnamefont {W.}~\bibnamefont {Belzig}},
  \bibinfo {author} {\bibfnamefont {T.}~\bibnamefont {Nussbaumer}}, \bibinfo
  {author} {\bibfnamefont {B.}~\bibnamefont {Babi\ifmmode~\acute{c}\else
  \'{c}\fi{}}}, \bibinfo {author} {\bibfnamefont {C.}~\bibnamefont {Bruder}},\
  and\ \bibinfo {author} {\bibfnamefont {C.}~\bibnamefont {Sch\"onenberger}},\
  }\bibfield  {title} {\bibinfo {title} {Multiple {Andreev} {R}eflections in a
  {C}arbon {N}anotube {Q}uantum {D}ot},\ }\href
  {https://doi.org/10.1103/PhysRevLett.91.057005} {\bibfield  {journal}
  {\bibinfo  {journal} {Phys. Rev. Lett.}\ }\textbf {\bibinfo {volume} {91}},\
  \bibinfo {pages} {057005} (\bibinfo {year} {2003})}\BibitemShut {NoStop}%
\bibitem [{\citenamefont {Zazunov}\ \emph {et~al.}(2006)\citenamefont
  {Zazunov}, \citenamefont {Egger}, \citenamefont {Mora},\ and\ \citenamefont
  {Martin}}]{Zazunov2006}%
  \BibitemOpen
  \bibfield  {author} {\bibinfo {author} {\bibfnamefont {A.}~\bibnamefont
  {Zazunov}}, \bibinfo {author} {\bibfnamefont {R.}~\bibnamefont {Egger}},
  \bibinfo {author} {\bibfnamefont {C.}~\bibnamefont {Mora}},\ and\ \bibinfo
  {author} {\bibfnamefont {T.}~\bibnamefont {Martin}},\ }\bibfield  {title}
  {\bibinfo {title} {Superconducting transport through a vibrating molecule},\
  }\href {https://doi.org/10.1103/PhysRevB.73.214501} {\bibfield  {journal}
  {\bibinfo  {journal} {Phys. Rev. B}\ }\textbf {\bibinfo {volume} {73}},\
  \bibinfo {pages} {214501} (\bibinfo {year} {2006})}\BibitemShut {NoStop}%
\bibitem [{\citenamefont {Lu}\ \emph {et~al.}(2020)\citenamefont {Lu},
  \citenamefont {Burset},\ and\ \citenamefont {Tanaka}}]{Lu2020}%
  \BibitemOpen
  \bibfield  {author} {\bibinfo {author} {\bibfnamefont {B.}~\bibnamefont
  {Lu}}, \bibinfo {author} {\bibfnamefont {P.}~\bibnamefont {Burset}},\ and\
  \bibinfo {author} {\bibfnamefont {Y.}~\bibnamefont {Tanaka}},\ }\bibfield
  {title} {\bibinfo {title} {Spin-polarized multiple {Andreev} reflections in
  spin-split superconductors},\ }\href
  {https://doi.org/10.1103/PhysRevB.101.020502} {\bibfield  {journal} {\bibinfo
   {journal} {Phys. Rev. B}\ }\textbf {\bibinfo {volume} {101}},\ \bibinfo
  {pages} {020502(R)} (\bibinfo {year} {2020})}\BibitemShut {NoStop}%
\bibitem [{\citenamefont {Badiane}\ \emph {et~al.}(2011)\citenamefont
  {Badiane}, \citenamefont {Houzet},\ and\ \citenamefont {Meyer}}]{Meyer2011}%
  \BibitemOpen
  \bibfield  {author} {\bibinfo {author} {\bibfnamefont {D.~M.}\ \bibnamefont
  {Badiane}}, \bibinfo {author} {\bibfnamefont {M.}~\bibnamefont {Houzet}},\
  and\ \bibinfo {author} {\bibfnamefont {J.~S.}\ \bibnamefont {Meyer}},\
  }\bibfield  {title} {\bibinfo {title} {Nonequilibrium {Josephson} {E}ffect
  through {H}elical {E}dge {S}tates},\ }\href
  {https://doi.org/10.1103/PhysRevLett.107.177002} {\bibfield  {journal}
  {\bibinfo  {journal} {Phys. Rev. Lett.}\ }\textbf {\bibinfo {volume} {107}},\
  \bibinfo {pages} {177002} (\bibinfo {year} {2011})}\BibitemShut {NoStop}%
\bibitem [{\citenamefont {San-Jose}\ \emph {et~al.}(2013)\citenamefont
  {San-Jose}, \citenamefont {Cayao}, \citenamefont {Prada},\ and\ \citenamefont
  {Aguado}}]{SanJose2013}%
  \BibitemOpen
  \bibfield  {author} {\bibinfo {author} {\bibfnamefont {P.}~\bibnamefont
  {San-Jose}}, \bibinfo {author} {\bibfnamefont {J.}~\bibnamefont {Cayao}},
  \bibinfo {author} {\bibfnamefont {E.}~\bibnamefont {Prada}},\ and\ \bibinfo
  {author} {\bibfnamefont {R.}~\bibnamefont {Aguado}},\ }\bibfield  {title}
  {\bibinfo {title} {Multiple {Andreev} reflection and critical current in
  topological superconducting nanowire junctions},\ }\href
  {https://doi.org/10.1088/1367-2630/15/7/075019} {\bibfield  {journal}
  {\bibinfo  {journal} {New Journal of Physics}\ }\textbf {\bibinfo {volume}
  {15}},\ \bibinfo {pages} {075019} (\bibinfo {year} {2013})}\BibitemShut
  {NoStop}%
\bibitem [{\citenamefont {Huang}\ \emph {et~al.}(2014)\citenamefont {Huang},
  \citenamefont {Leijnse}, \citenamefont {Flensberg},\ and\ \citenamefont
  {Xu}}]{Huang2014}%
  \BibitemOpen
  \bibfield  {author} {\bibinfo {author} {\bibfnamefont {G.-Y.}\ \bibnamefont
  {Huang}}, \bibinfo {author} {\bibfnamefont {M.}~\bibnamefont {Leijnse}},
  \bibinfo {author} {\bibfnamefont {K.}~\bibnamefont {Flensberg}},\ and\
  \bibinfo {author} {\bibfnamefont {H.~Q.}\ \bibnamefont {Xu}},\ }\bibfield
  {title} {\bibinfo {title} {Tunnel spectroscopy of {M}ajorana bound states in
  topological superconductor/quantum dot {J}osephson junctions},\ }\href
  {https://doi.org/10.1103/PhysRevB.90.214507} {\bibfield  {journal} {\bibinfo
  {journal} {Phys. Rev. B}\ }\textbf {\bibinfo {volume} {90}},\ \bibinfo
  {pages} {214507} (\bibinfo {year} {2014})}\BibitemShut {NoStop}%
\bibitem [{\citenamefont {Datta}(1995)}]{Datta1995}%
  \BibitemOpen
  \bibfield  {author} {\bibinfo {author} {\bibfnamefont {S.}~\bibnamefont
  {Datta}},\ }\href@noop {} {\emph {\bibinfo {title} {Electronic Transport in
  Mesoscopic Systems}}}\ (\bibinfo  {publisher} {Cambridge University Press},\
  \bibinfo {address} {Cambridge},\ \bibinfo {year} {1995})\BibitemShut
  {NoStop}%
\bibitem [{\citenamefont {Blonder}\ \emph {et~al.}(1982)\citenamefont
  {Blonder}, \citenamefont {Tinkham},\ and\ \citenamefont
  {Klapwijk}}]{Blonder1982}%
  \BibitemOpen
  \bibfield  {author} {\bibinfo {author} {\bibfnamefont {G.~E.}\ \bibnamefont
  {Blonder}}, \bibinfo {author} {\bibfnamefont {M.}~\bibnamefont {Tinkham}},\
  and\ \bibinfo {author} {\bibfnamefont {T.~M.}\ \bibnamefont {Klapwijk}},\
  }\bibfield  {title} {\bibinfo {title} {Transition from metallic to tunneling
  regimes in superconducting microconstrictions: Excess current, charge
  imbalance, and supercurrent conversion},\ }\href
  {https://doi.org/10.1103/PhysRevB.25.4515} {\bibfield  {journal} {\bibinfo
  {journal} {Phys. Rev. B}\ }\textbf {\bibinfo {volume} {25}},\ \bibinfo
  {pages} {4515} (\bibinfo {year} {1982})}\BibitemShut {NoStop}%
\bibitem [{\citenamefont {Martinez}(2003)}]{Martinez2003}%
  \BibitemOpen
  \bibfield  {author} {\bibinfo {author} {\bibfnamefont {D.~F.}\ \bibnamefont
  {Martinez}},\ }\bibfield  {title} {\bibinfo {title}
  {Floquet{\textendash}{Green} function formalism for harmonically driven
  {Hamiltonians}},\ }\href {https://doi.org/10.1088/0305-4470/36/38/302}
  {\bibfield  {journal} {\bibinfo  {journal} {Journal of Physics A:
  Mathematical and General}\ }\textbf {\bibinfo {volume} {36}},\ \bibinfo
  {pages} {9827} (\bibinfo {year} {2003})}\BibitemShut {NoStop}%
\bibitem [{\citenamefont {Tsuji}\ \emph {et~al.}(2008)\citenamefont {Tsuji},
  \citenamefont {Oka},\ and\ \citenamefont {Aoki}}]{Tsuji2008}%
  \BibitemOpen
  \bibfield  {author} {\bibinfo {author} {\bibfnamefont {N.}~\bibnamefont
  {Tsuji}}, \bibinfo {author} {\bibfnamefont {T.}~\bibnamefont {Oka}},\ and\
  \bibinfo {author} {\bibfnamefont {H.}~\bibnamefont {Aoki}},\ }\bibfield
  {title} {\bibinfo {title} {Correlated electron systems periodically driven
  out of equilibrium: $\text{Floquet}+\text{DMFT}$ formalism},\ }\href
  {https://doi.org/10.1103/PhysRevB.78.235124} {\bibfield  {journal} {\bibinfo
  {journal} {Phys. Rev. B}\ }\textbf {\bibinfo {volume} {78}},\ \bibinfo
  {pages} {235124} (\bibinfo {year} {2008})}\BibitemShut {NoStop}%
\bibitem [{\citenamefont {Haug}\ and\ \citenamefont {Jauho}(1996)}]{Jauho1996}%
  \BibitemOpen
  \bibfield  {author} {\bibinfo {author} {\bibfnamefont {H.}~\bibnamefont
  {Haug}}\ and\ \bibinfo {author} {\bibfnamefont {A.-P.}\ \bibnamefont
  {Jauho}},\ }\href@noop {} {\emph {\bibinfo {title} {Quantum Kinetics in
  Transport and Optics of Semiconductors}}}\ (\bibinfo  {publisher}
  {Springer-Verlag},\ \bibinfo {address} {Berlin},\ \bibinfo {year}
  {1996})\BibitemShut {NoStop}%
\bibitem [{\citenamefont {Draelos}\ \emph {et~al.}(2018)\citenamefont
  {Draelos}, \citenamefont {Wei}, \citenamefont {Seredinski}, \citenamefont
  {Ke}, \citenamefont {Mehta}, \citenamefont {Chamberlain}, \citenamefont
  {Watanabe}, \citenamefont {Taniguchi}, \citenamefont {Yamamoto},
  \citenamefont {Tarucha}, \citenamefont {Borzenets}, \citenamefont {Amet},\
  and\ \citenamefont {Finkelstein}}]{Draelos2018}%
  \BibitemOpen
  \bibfield  {author} {\bibinfo {author} {\bibfnamefont {A.~W.}\ \bibnamefont
  {Draelos}}, \bibinfo {author} {\bibfnamefont {M.~T.}\ \bibnamefont {Wei}},
  \bibinfo {author} {\bibfnamefont {A.}~\bibnamefont {Seredinski}}, \bibinfo
  {author} {\bibfnamefont {C.~T.}\ \bibnamefont {Ke}}, \bibinfo {author}
  {\bibfnamefont {Y.}~\bibnamefont {Mehta}}, \bibinfo {author} {\bibfnamefont
  {R.}~\bibnamefont {Chamberlain}}, \bibinfo {author} {\bibfnamefont
  {K.}~\bibnamefont {Watanabe}}, \bibinfo {author} {\bibfnamefont
  {T.}~\bibnamefont {Taniguchi}}, \bibinfo {author} {\bibfnamefont
  {M.}~\bibnamefont {Yamamoto}}, \bibinfo {author} {\bibfnamefont
  {S.}~\bibnamefont {Tarucha}}, \bibinfo {author} {\bibfnamefont {I.~V.}\
  \bibnamefont {Borzenets}}, \bibinfo {author} {\bibfnamefont {F.}~\bibnamefont
  {Amet}},\ and\ \bibinfo {author} {\bibfnamefont {G.}~\bibnamefont
  {Finkelstein}},\ }\bibfield  {title} {\bibinfo {title} {Investigation of
  supercurrent in the quantum {H}all regime in graphene {J}osephson
  junctions},\ }\href {https://doi.org/10.1007/s10909-018-1872-9} {\bibfield
  {journal} {\bibinfo  {journal} {Journal of Low Temperature Physics}\ }\textbf
  {\bibinfo {volume} {191}},\ \bibinfo {pages} {288} (\bibinfo {year}
  {2018})}\BibitemShut {NoStop}%
\bibitem [{\citenamefont {Bezuglyi}\ \emph {et~al.}(2017)\citenamefont
  {Bezuglyi}, \citenamefont {Bratus'},\ and\ \citenamefont
  {Shumeiko}}]{Bezuglyi2017}%
  \BibitemOpen
  \bibfield  {author} {\bibinfo {author} {\bibfnamefont {E.~V.}\ \bibnamefont
  {Bezuglyi}}, \bibinfo {author} {\bibfnamefont {E.~N.}\ \bibnamefont
  {Bratus'}},\ and\ \bibinfo {author} {\bibfnamefont {V.~S.}\ \bibnamefont
  {Shumeiko}},\ }\bibfield  {title} {\bibinfo {title} {Resonant subgap current
  transport in {Josephson} field effect transistor},\ }\href
  {https://doi.org/10.1103/PhysRevB.95.014522} {\bibfield  {journal} {\bibinfo
  {journal} {Phys. Rev. B}\ }\textbf {\bibinfo {volume} {95}},\ \bibinfo
  {pages} {014522} (\bibinfo {year} {2017})}\BibitemShut {NoStop}%
\bibitem [{\citenamefont {Johansson}\ \emph
  {et~al.}(1999{\natexlab{b}})\citenamefont {Johansson}, \citenamefont
  {Bratus'}, \citenamefont {Shumeiko},\ and\ \citenamefont
  {Wendin}}]{Johansson1999-bis}%
  \BibitemOpen
  \bibfield  {author} {\bibinfo {author} {\bibfnamefont {G.}~\bibnamefont
  {Johansson}}, \bibinfo {author} {\bibfnamefont {K.~N.}\ \bibnamefont
  {Bratus'}}, \bibinfo {author} {\bibfnamefont {V.}~\bibnamefont {Shumeiko}},\
  and\ \bibinfo {author} {\bibfnamefont {G.}~\bibnamefont {Wendin}},\
  }\bibfield  {title} {\bibinfo {title} {Multiple {Andreev} reflections as a
  transport problem in energy space},\ }\href
  {https://doi.org/10.1006/spmi.1999.0740} {\bibfield  {journal} {\bibinfo
  {journal} {Superlattices and Microstructures}\ }\textbf {\bibinfo {volume}
  {25}},\ \bibinfo {pages} {905} (\bibinfo {year}
  {1999}{\natexlab{b}})}\BibitemShut {NoStop}%
\bibitem [{\citenamefont {Dolcini}\ and\ \citenamefont
  {Dell'Anna}(2008)}]{Dolcini2008}%
  \BibitemOpen
  \bibfield  {author} {\bibinfo {author} {\bibfnamefont {F.}~\bibnamefont
  {Dolcini}}\ and\ \bibinfo {author} {\bibfnamefont {L.}~\bibnamefont
  {Dell'Anna}},\ }\bibfield  {title} {\bibinfo {title} {Multiple {Andreev}
  reflections in a quantum dot coupled to superconducting leads: {Effect} of
  spin-orbit coupling},\ }\href {https://doi.org/10.1103/PhysRevB.78.024518}
  {\bibfield  {journal} {\bibinfo  {journal} {Phys. Rev. B}\ }\textbf {\bibinfo
  {volume} {78}},\ \bibinfo {pages} {024518} (\bibinfo {year}
  {2008})}\BibitemShut {NoStop}%
\end{thebibliography}
\end{document}